%% file: luminous_red_novae.tex
\documentclass[a4paper,fleqn,usenatbib,useAMS,usedcolumn]{mnras}

\usepackage[T1]{fontenc}
\usepackage{ae,aecompl}
\usepackage[utf8]{inputenc}

\usepackage{xcolor}

\usepackage{graphicx} 
\DeclareGraphicsExtensions{.pdf,.png,.jpg,.ps}
\graphicspath{{../plots/}}

\usepackage{subcaption}
\usepackage{float}

\usepackage[fleqn]{amsmath} 
\usepackage{amssymb}  
\usepackage{gensymb}
\usepackage{hyperref}
\usepackage[nolist]{acronym}
\usepackage{ulem}
\usepackage{etoolbox}
\usepackage{booktabs}

\acrodef{GW}{gravitational-wave}
\acrodef{NS}{neutron star}
\acrodef{BH}{black hole}
\acrodef{BBH}{binary black hole}
\acrodef{aLIGO}{Advanced LIGO}
\acrodef{AdV}{Advanced Virgo}
\acrodef{PN}{post-Newtonian}
\acrodef{BNS}{binary neutron star}
\acrodef{MS}{main-sequence}
\acrodef{HG}{Hertzsprung-Gap}
\acrodef{CHeB}{core helium-burning}
\acrodef{HeMS}{helium star}
\acrodef{HeHG}{helium Hertzsprung-gap}
\acrodef{HeGB}{helium giant}
\acrodef{WR}{Wolf-Rayet}
\acrodef{PE}{parameter estimation}
\acrodef{SNR}{signal-to-noise ratio}
\acrodef{IMF}{initial mass function}
\acrodef{CE}{common-envelope}
\acrodef{MCMC}{Markov-chain Monte Carlo}
\acrodef{LBV}{Luminous Blue Variable}
\acrodef{COMPAS}{Compact Object Mergers: Population Astrophysics and Statistics}

\title[Luminous Red Novae]{Luminous Red Novae: population models and future prospects}

\author[]{\parbox{\textwidth}{
George Howitt$^{1,2,3}$\thanks{E-mail: ghowitt@student.unimelb.edu.au},
Simon Stevenson$^{4,3,2}$,
Alejandro Vigna-G\'{o}mez$^{6,5,3,2}$,
Stephen Justham$^{7,8,9}$,
Natasha Ivanova$^{10}$,
Tyrone E. Woods$^{2}$,
Coenraad J. Neijssel$^{5,3,2,11}$,
Ilya Mandel$^{5,3,2}$
}
\vspace{0.5cm}\\
\parbox{\textwidth}{$^{1}$ School of Physics, University of Melbourne, Parkville, Victoria, 3010, Australia  \\
$^{2}$ Institute for Gravitational Wave Astronomy and School of Physics and Astronomy, University of Birmingham, Edgbaston, Birmingham B15 2TT, United Kingdom \\
$^{3}$ OzGrav, Australian Research Council Centre of Excellence for Gravitational Wave Discovery \\
$^{4}$ Centre for Astrophysics and Supercomputing, Swinburne University of Technology, Hawthorn, 3122, Victoria, Australia \\
$^5$ Monash Centre for Astrophysics, School of Physics and Astronomy, Monash University, Clayton, Victoria 3800, Australia\\
$^6$ DARK, Niels Bohr Institute, University of Copenhagen, Blegdamsvej 17, 2100, Copenhagen, Denmark\\
$^{7}$ School of Astronomy \& Space Science, University of the Chinese Academy of Sciences, Beijing, China \\
$^{8}$ National Astronomical Observatories, Chinese Academy of Sciences, Beijing 100012, China \\
$^{9}$Astronomical Institute Anton Pannekoek, University of Amsterdam, P.O. Box 94249, 1090 GE, Amsterdam, The Netherlands \\
$^{10}$ Department of Physics, University of Alberta, 11322-89 Ave, Edmonton, AB, T6G2E7, Canada} \\
$^{11}$ Max Planck Institute for Gravitational Physics (Albert Einstein Institute), D-30167 Hannover, Germany
}
\date{\today}

\pubyear{2018}

\begin{document}
\label{firstpage}
\pagerange{\pageref{firstpage}--\pageref{lastpage}}
\maketitle

\begin{abstract}
A class of optical transients known as Luminous Red Novae (LRNe) have recently been associated with mass ejections from binary stars undergoing common-envelope evolution.
We use the population synthesis code COMPAS to explore the impact of a range of assumptions about the physics of common-envelope evolution on the properties of LRNe.
In particular, we investigate the influence of various models for the energetics of LRNe on the expected event rate and light curve characteristics, and compare with the existing sample.
We find that the Galactic rate of LRNe is $\sim 0.2$ yr$^{-1}$, in agreement with the observed rate. 
In our models, the luminosity function of Galactic LRNe covers multiple decades in luminosity and is dominated by signals from stellar mergers, consistent with observational constraints from iPTF and the Galactic sample of LRNe.
We discuss how observations of the brightest LRNe may provide indirect evidence for the existence of massive ($> 40$ M$_\odot$) red supergiants.  Such LRNe could be markers along the evolutionary pathway leading to the formation of double compact objects.
We make predictions for the population of LRNe observable in future transient surveys with the Large Synoptic Survey Telescope and the Zwicky Transient Facility. 
In all plausible circumstances, we predict a selection-limited observable population dominated by bright, long-duration events caused by common envelope ejections.
We show that the Large Synoptic Survey Telescope will observe $20$--$750$ LRNe per year, quickly constraining the luminosity function of LRNe and probing the physics of common-envelope events.
\end{abstract}




\section{Introduction}
\label{sec:Introduction}

Common-envelope evolution
\citep{Paczynski:1976} 
is a phase of mass transfer in the evolution of many stellar binaries, wherein the two stars or stellar cores orbit inside a shared gas envelope. The resulting drag force causes significant energy dissipation and a rapid decay of the binary's orbit.
The outcome is either a stellar merger or, if the envelope is successfully ejected, a binary with a much-reduced separation.
The common-envelope phase is thought to be an important evolutionary channel for the formation of X-ray binaries, binary pulsars, and gravitational-wave sources such as merging double white dwarfs, binary neutron stars, and double black holes 
\citep[e.g.][]{SmarrBlandford:1976, vdH:1976, Tutukov:1993,Voss:2003ep,Dominik:2012kk,Belczynski:2016obo,Stevenson:2017tfq,Vignagomez2018}.

The short duration of the common envelope phase makes catching a binary during this process observationally challenging. Recently, a class of optical transients in the luminosity gap between novae and supernovae known as Luminous Red Novae (LRNe) have been associated with common-envelope evolution 
\citep{SokerTylenda:2003,Kulkarni:2007kf,Tylenda:2011,Ivanova:2013db,Pastorello2019}.
These transients are characterised by a rapid rise in luminosity followed by a lengthy plateau, often with a secondary maximum which may be powered by the shock collision of the expanding shell with previously ejected material \citep{Metzger:2017wrz}.
The best evidence for the association of LRNe with common-envelope evolution comes from the observed orbital decay of the Galactic binary V1309 Sco over ten years 
\citep{Tylenda:2011} 
preceeding a LRN outburst. 
\citet{SokerTylenda:2003}, \citet{Kulkarni:2007kf}, \citet{Tylenda:2011}, and \citet{Williams:2015}
have suggested that LRNe may be due to stellar mergers, while 
\citet{Blagorodnova:2017} 
proposed that the LRN M101 OT2015-1 was produced by the ejection of a common envelope from a binary involving a massive star. 
\citet{Pastorello2019} presented photometric and spectroscopic data on several bright LRNe and LRN candidates, finding that the ejection of a common-envelope was the most favoured origin of these transients.

\input{observations_table}

To date, there have been 13 observations of LRNe with recorded plateau durations and accurate distance estimates allowing inference on the absolute magnitude.
We summarise these in Table~\ref{tab:observations}, including the characteristic luminosities and durations of the plateaux, as well as whether they are of Galactic or extra-galactic origin. 
Where plateau durations are not stated in the discovery papers, we estimate them from the light curves as the time over which the brightness decreases by approximately one magnitude (either $V$-band where given or absolute) from the peak.
In the case of V838 Mon, NGC 4490-OT2011, and M101 OT2015, which have a double-peaked lightcurve, we use the time between the first and second maxima to estimate the duration.
All luminosities quoted in table \ref{tab:observations} are taken from the original references cited, or converted from bolometric magnitudes, with the exception of M85--OT, where we use the luminosity in the \textit{R} band from 
\citet{Kulkarni:2007kf}.
Error ranges are approximate and are determined by eye.
Some events described as possible LRNe in the literature, such as V4332 Sgr 
\citep{Martini1999,Kimeswenger:2005bva}, 
have been omitted in our analysis; their absolute magnitude is not known, and so we cannot compare their properties to our simulated population of LRNe. There are other optical transients with similar luminosities to LRNe. 
These contribute a source of confusion noise to the population of LRNe with a possibly different physical mechanism behind the transient. 
We do not regard SN2008S \citep{2008CBET.1234....1A}, NGC 300 OT 2008 \citep{Berger2009,Prieto2009,Bond2009} and M51 OT2019-1 \citep{Jencson:2019ApJL} as LRNe, but rather as likely Type IIn electron capture supernovae \citep{Prieto2008,Adams:2015xjy}. 
We also neglect the luminous infra-red transient VVV-WIT-06, which is most likely an obscured classical nova \citep{Banerjee2018}. 
Nova Vul 1670 has been suggested as a possible historical example of a LRN \citep{2015Natur.520..322K}, however we do not include it in our present sample.
Due to its unusual light curve, we also exclude OGLE-2002-BLG-360 \citep{Tylenda:2013} from our subsequent analysis.
We exclude LRN PTF 10FQS from our sample; it is likely this is an intermediate-luminosity red transient, and not an LRN
\citep{Kasliwal2011,Pastorello2019}.
On the other hand, we include the transient UGC 12307--2013OT1 in our sample, despite the uncertainty over its classification, owing to its late discovery  \citep{Pastorello2019}; its inclusion in our sample does not affect the findings of this paper.
In the future, better models and larger statistical samples could make it possible to use tools such as those introduced by 
\citet{Farr:2015} 
for counting amid confusion to avoid making binary cuts on potential LRN candidates.

As Table \ref{tab:observations} shows, the current scarcity and diversity of LRN observations makes extracting information about the intrinsic population of these transients and their progenitors difficult, though some progress has been made. The unusually luminous extragalactic LRN NGC 4490-OT2011 has been identified with a merger involving a massive blue progenitor \citep{Smith:2016qtr}.
Within our Galaxy, only four likely LRNe have been detected in approximately 25 years: V4332 Sgr, V1309 Sco, V838 Mon, and tentatively OGLE-2002-BLG-360 (though see above).
The Large Synoptic Survey Telescope (LSST) will observe $\sim$20,000 deg$^2$ with a cadence of \textbf{$\sim 3$} days down to a single-visit limiting $r$-band magnitude of $\approx 24.2$ \citep{LSST2017}, and is expected to increase the number of detected LRNe by several orders of magnitude \citep{LSSTScience:2009}.

In this paper, we use the binary population synthesis code COMPAS 
\citep{Stevenson:2017tfq,Barrett:2017FIM,Vignagomez2018,Neijssel:2019} 
and a model of LRN plateau durations and luminosities from 
\citet{Ivanova:2013db} 
to predict what the properties of this observed population will be under a variety of models of common envelope interaction.
We find that the Galactic rate of LRNe is $\approx 0.1$ yr$^{-1}$, consistent with observations and previous population studies.
We predict that the volumetric rate in the local Universe is $\approx 8 \times 10^{-4}$ Mpc$^{-3}$ yr$^{-1}$, and that with LSST the rate of LRN detections could be as high as 750 yr$^{-1}$. 

The rest of this paper is structured as follows. 
In section~\ref{subsec:pop_synth} we introduce the population synthesis code COMPAS, 
which we use to simulate a catalogue of common-envelope events.
In section~\ref{subsec:mass_transfer}, we describe how mass transfer, including common-envelope evolution, is implemented within COMPAS.
In section~\ref{subsec:simulation_parameters} we describe the parameters of the simulations used in this paper. 
In section~\ref{subsec:LRNe} we summarise the formalism that we employ for predicting the observable properties of LRNe. 
In section~\ref{subsec:pop_stats} we present the results of the population synthesis simulation and compare to previous work.
In sections~\ref{subsec:plateau} and \ref{subsec:event_rates} we present our predictions for the observable properties and rates of LRNe, respectively.  
We discuss these results in section~\ref{sec:discussion}. 

\section{Methods}
\label{sec:methods}

\subsection{Population synthesis}
\label{subsec:pop_synth}

In order to predict the rate and properties of common-envelope events, we simulate a large stellar population using the rapid population synthesis module of COMPAS
\citep{Stevenson:2017tfq,Barrett:2017FIM,Vignagomez2018,Neijssel:2019}.
The COMPAS binary population synthesis module simulates a population of binaries by Monte Carlo sampling a distribution of initial (ZAMS) binary masses and separations, then evolving each star in the binary according to the single-stellar evolution prescription of 
\citet{Hurley2000}.
COMPAS evolves each binary by modelling the physics of mass loss and transfer due to effects such as stellar winds, Roche lobe overflow and supernovae, until the binary either merges, becomes unbound, or forms a double compact object.
We describe in detail how mass transfer during binary evolution is implemented in COMPAS below.

The version of COMPAS used in this paper, and the parameters used in our simulation, are similar to those described in \citet{Vignagomez2018}. 
Where substantive changes have been made to either we describe these explicitly in text.

\subsection{The common envelope}
\label{subsec:mass_transfer}

In the COMPAS models, a Roche-lobe overflow mass transfer episode begins when the radius of one of the star becomes larger than the effective radius of its Roche lobe \citep{Eggleton1983rocheLobe}.  
This may occur due to radial expansion of a star, orbital evolution of the binary, or both. 

A mass transfer episode either ends with the system gently decoupling from Roche-lobe overflow, or mass loss from the donor leads to a runaway process. The first of these scenarios, dynamically stable mass transfer, occurs on either the nuclear or thermal timescale of the donor.  In the second scenario, dynamical instability leads to the formation of a common envelope, with subsequent inspiral on the dynamical timescale \citep{Paczynski:1976}. 

The stability of mass transfer is determined in COMPAS using the mass-radius relationship, $\zeta=d \log R/d \log M$, in order to quantify how the radius responds to mass loss
\citep{soberman1997stability,1997MNRAS.291..732T}. 
We compare the adiabatic mass-radius coefficient of the donor star $\zeta_{\rm{ad}}$ to the Roche-lobe mass-radius coefficient of the binary $\zeta_{\rm{RL}}$. The latter is computed under the assumption of stable mass transfer, with the amount of angular momentum lost from the system given by the fiducial COMPAS model, as described in section 2.3.1 of \citet{Neijssel:2019}. In order for the mass transfer episode to be dynamically stable, we require $\zeta_{\rm{ad}} \ge \zeta_{\rm{RL}}$.   
If this condition is not satisfied, mass transfer is assumed to be unstable on a dynamical timescale and leads to a common-envelope phase.    We use the choices and implementation of $\zeta_{\rm{ad}}$ from \cite{Vignagomez2018}, informed in part by quantitative comparisons to observations of Galactic double neutron stars and merging binary black holes.  In addition to the standard picture of a common envelope described above, 
these choices are intended to phenomenologically account for common envelopes initiated by the expansion of the accretor beyond its Roche lobe in response to rapid mass transfer \citep{Nariai:1976}.

For \ac{MS} and \ac{HG} donors, we use fixed values of $\zeta_{\rm{ad,MS}}=2.0$ and $\zeta_{\rm{ad,HG}}=6.5$, respectively. These mass-radius relations represent roughly average typical values for these phases, following adiabatic mass-loss models from \cite{ge2015adiabatic}.  
The prescription for \ac{HG} donors is an effective value intended to describe delayed dynamical instability \citep{Hjellming:1987}, rather than incipient Roche-lobe overflow from a radiative \ac{HG} donor.
For core-helium burning and giant stars we use a fit for condensed polytropes with convective envelopes presented  in \cite{soberman1997stability}, $\zeta_{\rm{ad}}=\zeta_{\rm{SPH}}$. Finally, for stripped stars, such as naked helium, helium-shell-burning or helium giant stars, we assume mass transfer to always be stable; this results in having no LRNe from these stripped donors in our model.

\subsubsection{Common-envelope evolution}
\label{subsubsec:common_envelope}

The common-envelope phase remains a poorly understood key aspect of binary evolution.   We use the standard assumption that if $\zeta_{\rm{ad}} < \zeta_{\rm{RL}}$, mass transfer is unstable and leads to a common-envelope phase \citep{Paczynski:1976}.  However, the stability threshold is uncertain: for example, apparent initial instability for \ac{HG} donors could resolve into stable mass transfer \citep{Pavlovskii:2016}.  

It is predicted to lead to a significant tightening of the binary, and thus contribute to both stellar mergers and double compact object formation \citep[see, e.g., discussion in ][]{Ivanova:2012vx}. 
The common-envelope phase is initiated when unstable mass transfer allows the envelope of the donor to engulf the companion as well as the core of the donor, so that both stars/stellar cores orbit inside the recently formed common envelope. 
The binary formed by the companion and the donor's core is not in co-rotation with the envelope. 
Viscous shear and tidal interactions with the envelope cause the companion to inspiral.
This inspiral liberates orbital energy.
If the released energy is large enough, and is effectively transferred to the envelope, the envelope may be ejected; otherwise, the system merges.
The conditions that are assumed to lead to either envelope ejection or stellar merger within the envelope are discussed below.

\subsubsection{Energy formalism}
\label{subsubsec:energy_formalism}

In the classic energy $\alpha$-formalism for the common envelope  
\citep{Iben:1984,Webbink:1984,Webbink:2008,DeMarco2011,Ivanova:2012vx}, the change in orbital energy during the phase is compared to the binding energy required to eject the envelope to infinity.
The difference in the initial and final orbital energies is given by
\begin{equation}
\Delta E_\mathrm{orb} =  E_\mathrm{initial} - E_\mathrm{final} = \left( - \frac{G M_{1} M_{2}}{2a_\mathrm{initial}} +  \frac{G M_{1,c} M_{2}}{2a_\mathrm{final}} \right) , 
\label{eq:delta_Eorb_ce}
\end{equation}
where $a_\mathrm{initial}$ and $a_\mathrm{final}$ are the initial and final orbital separations respectively, $M_{1}$ and $M_{2}$ are the initial masses of the two stars and $M_{1,c}$ is the core mass of the donor after its envelope is removed. 
The parameter $\alpha$ characterises the efficiency with which the orbital energy is used to eject the envelope whose initial binding energy is $E_\mathrm{bind}$:
\begin{equation}
\alpha \Delta E_\mathrm{orb} = E_\mathrm{bind} \, .
\label{eq:alpha_ce}
\end{equation}
If the binding and orbital energies are the only energies involved in the common envelope interaction, then $0 \leq \alpha \leq 1$.
There are, however, additional possible energy sources not taken into account in the classical definition, such as recombination energy \citep{Nandez2016} 
or enthalpy 
\citep{Ivanova2011a}, 
which allow for $\alpha > 1$.
Other recent work suggests that $\alpha < 0.6$--1.0 \citep{Iaconi2019}.
Generally speaking, a higher value of $\alpha$ leads to more binaries surviving the common-envelope phase, while a lower value leads to more mergers during common-envelope evolution.
In this paper, we assume $\alpha = 1$, although this assumption technically violates energy conservation in some of the models for ejecta kinetic energy described below.
To check the impact of the assumed value of $\alpha$ on our results, we repeat a population synthesis simulation with $\alpha=0.5$.
We discuss the effects of modifying this assumption in section \ref{subsec:pop_stats}.

The binding energy of the envelope $E_\mathrm{bind}$ can be calculated from detailed stellar models for single stars 
\cite[e.g.][]{DewiTauris:2000,Xu2010,Loveridge2011,Ivanova2011b,Wang2016,Kruckow2016}.
The main source of uncertainty in calculating the binding energy is in determining the core-envelope boundary 
\citep[e.g.][]{Tauris:2001cx,Ivanova2011b}. 
In COMPAS, we use the parameter $\lambda$ 
\citep{deKool:1987}
to characterise the binding energy of a stellar envelope, so we can re-write Equations~\ref{eq:delta_Eorb_ce} and \ref{eq:alpha_ce} as \citep{Ivanova:2012vx}
\begin{equation}
\frac{G M_1M_{1,\mathrm{env}}}{\lambda R_1} = 
\alpha \left(-\frac{G M_1 M_2}{2 a_{\mathrm{initial}}} + \frac{G M_{1,\mathrm{c}}M_2}{2 a_{\mathrm{final}}} \right)\, ,
\label{energy_equation}
\end{equation}
where $R_1$ is the radius of the donor before the interaction.
Equation~\ref{energy_equation} assumes that only the donor star has a core-envelope separation. 
If the companion also has an envelope, then we assume that unstable mass transfer from either star triggers a `double-core common envelope', in which case Equation~\ref{energy_equation} becomes
\begin{equation}
\frac{M_1M_{1,\mathrm{env}}}{\lambda_1 R_1} + \frac{M_2M_{2,\mathrm{env}}}{\lambda_2 R_2}= 
\alpha \left(-\frac{G M_1 M_2}{2 a_{\mathrm{initial}}} + \frac{G M_{1,\mathrm{c}}M_{2,\mathrm{c}}}{2 a_{\mathrm{final}}} \right)\, .
\label{energy_equation_double_CE}
\end{equation}
In this work, we follow 
\citet{Vignagomez2018}
in computing $\lambda$ using fitting formulae to the detailed stellar structure models of 
\citet{Xu2010,XuLi:2010err}\footnote{We use the $\lambda_b$ values from \citet{Xu2010,XuLi:2010err},  which include the internal energy terms.}. 
In order to determine whether a common envelope interaction results in a merger or the ejection of the envelope, we solve Equation \ref{energy_equation} or \ref{energy_equation_double_CE}  for $a_\mathrm{final}$.
If both stars or stellar cores fit within their Roche lobes at the end of the common-envelope phase, we assume that the envelope has been ejected; otherwise, we assume that the stars merge.

\subsubsection{Common-envelope evolution in COMPAS}
\label{subsubsec:ce_in_compas}

COMPAS does not directly simulate the common-envelope interaction -- such simulations are difficult and expensive even for single systems -- and so our implementation contains several prescriptive assumptions for how common-envelope evolution proceeds. 
Here we list several key assumptions that may differ between this work and other population synthesis studies.
\begin{itemize}
	\item  If the donor during a common-envelope phase is a \ac{MS} star, the result is always a stellar merger. 
	\item  The main change from the version of COMPAS used for \citet{Vignagomez2018}
	is that \ac{MS} accretors are allowed to engage in and survive a common-envelope phase.  Previously, dynamically unstable mass transfer onto \ac{MS} accretors was assumed to inevitably lead to mergers, which is still the case for \ac{MS} donors. This change leads to a higher rate of common-envelope ejections compared to the previous model.
	\item Some studies suggest that \ac{HG} stars do not have a clear core/envelope separation, so unstable mass transfer involving \ac{HG} donors should always lead to mergers \citep{Belczynski:2006zi,Dominik:2012kk}, similarly to \ac{MS} donors. We flag these systems and follow their evolution. If we assume that a \ac{HG} donor has a clear core/envelope separation, we evolve the system  by following the common-envelope energy formalism described above, in what is referred to as the ``optimistic'' variant. Alternatively, the system always merges in the ``pessimistic'' variant. We examine the impact of these variants on our results in section \ref{sec:results}. 
	\item If the stripped core overflows its Roche lobe immediately after the common envelope is ejected, we assume that the binary does not successfully emerge from the common-envelope phase and the stars merge.
\end{itemize}

\subsection{Simulation parameters}
\label{subsec:simulation_parameters}

For our model population, we evolve $5 \times 10^5$ binary systems. 
We draw the mass of the primary from the Kroupa \ac{IMF} \citep{Kroupa:2000iv}, with $1.0 M_\odot \leq M_1 \leq 100 M_\odot$. 
The lower mass limit is chosen so that we only simulate systems in which the primary will evolve off the main sequence within approximately the age of the Universe.
We neglect the effect of magnetic braking in low mass stars.
The upper mass limit is chosen due to the uncertainty of stellar evolution models in the high mass range.
We take our ZAMS mass cut into account when we normalise our simulated event rates by the total star formation rate.
The secondary mass is determined by drawing a mass ratio from a flat distribution \citep{Sana:2012}, with a minimum secondary mass of $0.1 M_\odot$, corresponding to the brown dwarf limit \citep{1963ApJ...137.1121K,1963PThPh..30..460H}.
The initial binary separation is drawn from a flat-in-the-log distribution with $0.01\, \mathrm{AU} \leq a \leq 1000\, \mathrm{AU}$ \citep{Abt:1983}.
While it is possible that binary systems may be born with separations as wide as $10^5 \mathrm{AU}$, these are not expected to interact and can be accounted for through normalisation.  In fact, only around half of the systems we simulate undergo any form of mass transfer within a Hubble time. 

\citet{Moe:2017ApJS} inferred a more complex distribution of binary initial conditions from observations, with correlations between component masses, orbital separations, and birth eccentricities.  
\citet{Klencki:2018} compared the impact of the \citet{Moe:2017ApJS} initial conditions against initial conditions similar to the ones we assumed here on the merger rate of double compact objects, and found that the effect of changing the initial distribution was generally well within other modelling uncertainties.  
Therefore, we opt for the simpler, non-correlated initial conditions in this study, but caution that this is one of several sources of uncertainty.
 
The initial conditions used in this simulation
We use a global ``solar'' metallicity of $Z = 0.0142$
\citep{Asplund:2009}.
In this work, we assume a continuous, constant-rate star formation of infinite duration; we discuss the effect of this assumption and the binary fraction/separation completeness further in section~\ref{sec:discussion}.

\subsection{Luminous red novae}
\label{subsec:LRNe}

To determine the observational properties of LRNe resulting from common envelope interactions in our population synthesis simulations, we follow 
\citet{Ivanova:2013db},
and adapt the scaling relations of
\citet{Popov1993} and \citet{Kasen2009},
derived for type IIP supernovae, to estimate the luminosity and duration of the LRN plateau:
%
\begin{multline}
L_p = 1.7 \times 10^4 L_\odot 
\left( \frac{R_\mathrm{init}}{3.5 \, R	_\odot} \right) ^{2/3}
\left( \frac{E_\mathrm{k}^\infty}{10^{46}  \, \mathrm{erg}} \right) ^{5/6} \\
\left( \frac{M_\mathrm{unb}}{0.03 \, M_\odot} \right) ^{-1/2} 
\left( \frac{\kappa}{0.32 \, \mathrm{cm}^{2} \, \mathrm{g}^{-1}} \right) ^{-1/3}
\left( \frac{T_\mathrm{rec}}{ 4500 \, \mathrm{K}} \right) ^{4/3} ,
\label{eq:plateau_luminosity}
\end{multline}
\begin{multline}
t_p = 17 \, \mathrm{days} 
\left( \frac{R_\mathrm{init}}{3.5 \, R_\odot} \right) ^{1/6}
\left( \frac{E_\mathrm{k}^\infty}{10^{46}  \, \mathrm{erg}} \right) ^{-1/6} \\
\left( \frac{M_\mathrm{unb}}{0.03 \, M_\odot} \right) ^{1/2} 
\left( \frac{\kappa}{0.32 \, \mathrm{cm}^{2} \, \mathrm{g}^{-1}} \right) ^{1/6}
\left( \frac{T_\mathrm{rec}}{ 4500 \, \mathrm{K}} \right) ^{-2/3} ,
\label{eq:plateau_duration}
\end{multline}
where $R_\mathrm{init}$ is the Roche lobe radius of the donor star prior to the common envelope phase (or the binary separation in the case of double-core common envelope events),
$E_\mathrm{k}^\infty$ is the kinetic energy of the ejected material after it escapes the gravitational potential well,
$M_\mathrm{unb}$ is the mass of the ejected material,
$\kappa$ is the opacity of the ionized ejecta,
and $T_\mathrm{rec}$ is the recombination temperature.
  
In this work, we use as fiducial values $\kappa = 0.32 \, \mathrm{cm}^{2} \, \mathrm{g}^{-1}$ and $T_\mathrm{rec} = 4500 \, \mathrm{K}$.
We parametrize $M_\mathrm{unb} = f_\mathrm{m} M_\mathrm{env}$, where 
$0 \leq f_\mathrm{m} \leq 1$ and $M_\mathrm{env}$ is the mass of the envelope.
We assume $f_\mathrm{m} = 1$ for successful common envelope ejection, corresponding to the total expulsion of the envelope.  We explore the effect of varying values of $f_\mathrm{m}$ for common envelope interactions leading to stellar mergers in section~\ref{sec:results}, where we consider $f_\mathrm{m}=0.05$ and $f_\mathrm{m}=0.5$.\footnote{\citet{Segev:2019} argue that common-envelope interactions of cool giants lead to a merger while ejecting the entire envelope.}
Some binaries ($\approx 6\%$) in our simulation undergo merger while both stars are still on the main sequence, before either has developed a core/envelope separation.
Simulations of stellar mergers by direct collision find that $\approx 1-10 \%$ of the total mass may be lost 
\citep{Lombardi2002,Glebbeek2013}.
Simulations of stellar mergers due to unstable mass transfer in binaries have also shown that a few percent of the primary mass is ejected
\citep{Nandez2014}.
We model MS-MS mergers in the same way as mergers resulting from a common-envelope interaction, but taking $M_\mathrm{env}$ to be 25$\%$ of the total mass of the binary; for $f_\mathrm{m}=0.05$ (0.5), this corresponds to 1.25\% (12.5\%) of the total MS-MS binary's mass being ejected during a merger, so the two variants we consider likely bracket the true value.

\subsection{Ejecta kinetic energy}
\label{subsec:ejecta_kinetic_energy}

The kinetic energy of the ejecta $E_\mathrm{k}^\infty$ is the most poorly determined quantity in equations \ref{eq:plateau_luminosity} and \ref{eq:plateau_duration}. 
We consider several prescriptions for calculating $E_\mathrm{k}^\infty$:
\begin{enumerate}
\item As the default model, we use the original prescription from 
\cite{Ivanova:2013db}, who assume that the kinetic energy of the ejecta
is proportional to the gravitational potential energy of the ejected material at the donor's surface before the interaction,
\begin{equation}
E_\mathrm{k}^\infty = \zeta(GM_\mathrm{unb} M)/R_\mathrm{init} \, ,
\end{equation}
\label{eq:ek_inf_original}
where $M$ is the mass of the donor, or, for a double-core common-envelope, the total mass in the binary.
\cite{Ivanova:2013db} consider several values of $\zeta$; here we use $\zeta = 10$ for ejections and $\zeta = 1$ for mergers. This implies ejecting the material with $v_\infty = \sqrt{\zeta} v_{\mathrm{esc,don}}$, where $v_{\mathrm{esc,don}}$ is the escape velocity from the surface of the donor.  If we enforce conservation of energy, $\Delta E_\mathrm{orb} = E_\mathrm{k}^\infty+ E_\mathrm{bind}$, this variant corresponds to a variable $\alpha$ given by $\alpha = 1/(1+\zeta \lambda)$, although we use a fixed $\alpha=1$ to determine whether the binary is able to eject the envelope in all variants for consistency.
\item We assume that the velocity of the ejected material is the same as the escape velocity from the binary, so that 
\begin{equation}
E_\mathrm{k}^\infty = \frac{1}{2}M_{\mathrm{unb}} v_{\mathrm{esc,bin}}^2 
= \frac{G M_{\mathrm{unb}} (M_1 + M_2 - M_{\mathrm{unb}})}{a} \, ,
\label{eq:ek_inf_escape}
\end{equation}
where 
\begin{enumerate}
\item $a=a_i$ for both mergers and ejections (the pre-CE escape velocity prescription), or,
\item $a=a_i$ for mergers and $a=a_f$ for ejections (the post-CE escape velocity prescription).
\end{enumerate}
These two variants span a broad range of ejecta kinetic energies in units of the envelope binding energy depending primarily on the mass ratio at the moment of CE onset, with significantly reduced ejecta energies for (ii-a) relatively to (ii-b) when the envelope is ejected and the binary hardens. 
\item Alternatively, we use a prescription based on the simulations presented in \cite{Nandez2016}, calibrated to few-solar-mass systems, where for mergers we use the same value for $E_\mathrm{k}^\infty$ as in prescriptions (ii-a) and (ii-b), and for ejections we use
\begin{equation}
E_\mathrm{k}^\infty = 0.3 \Delta E_\mathrm{orb}\, .
\end{equation}
\label{eq:ek_inf_nandez2016}
Variant (iii) corresponds to energy conservation with a fixed $\alpha=0.7$.  
\end{enumerate}

\subsection{Selection effects}
\label{subsec:selection effects}

In order to compute the detection rates and properties of observable LRNe, we need to take survey selection effects into account.   
For each simulated LRN we compute the maximum cosmological volume in which the event can be detected in a magnitude-limited survey, while making the simplifying assumption of a static Universe.
We then compute the star formation rate within the detectable volume of each event.
For distances $D <100$ Mpc, we calculate the star formation rate from the blue luminosity. 
We take the cumulative $B$-band luminosity to distance $D$ from the Gravitational-Wave Galaxy Catalogue \citep{White2011}, 
and convert from luminosity to star formation rate using the approximate Milky Way values of $2 M_\odot$ yr$^{-1}$ of star formation with a $B$-band luminosity of $1.5 \times 10^{10} L_\odot$
\citep{Licquia2015a,Licquia2015b}.
Beyond 100 Mpc, where the galaxy catalogue is incomplete\footnote{The GWGC is not complete even for $D < 100$ Mpc
\citep{Kulkarni2018}.
However, since the majority of our detections are expected to come from sources at greater than 100 Mpc, this does not affect our results.}, we assume a global star formation rate of $0.015 M_\odot$Mpc$^{-3}$yr$^{-1}$ 
\citep{Madau2014}.
The contribution of each simulated event to the event rate is the ratio of the integrated star formation rate in the observable volume of the event to the total evolved stellar mass represented by our simulation.

We consider the LSST limiting magnitude to be the median $r$-band value for the Wide, Fast, Deep survey of $24.16$ \citep{LSST2017}.  
The LSST Wide, Fast, Deep Survey will monitor 18000 square degrees with an average time between visits of 3 days. 
We account for the observed sky fraction in our predicted detection rates, but not for their duration: since more than 99\% of LRNe last for longer than 10 days in our fiducial model, we do not place any further cuts on plateau duration.  
We also predict LRN detection rates for the Zwicky Transient Facility (ZTF), which has a similar sky coverage and observing strategy to LSST, but with an average single-visit limiting $r$-band magnitude of $20.6$
\citep{Bellm:2019PASP}.
In applying selection effects, we do not apply a bolometric correction when converting between LRN luminosities calculated from equation \eqref{eq:plateau_luminosity} and magnitudes.
The LRN SED is presently poorly constrained.  Our model explicitly assumes a universal value of $T_{\rm rec} = 4500$ K.
As the population of detected LRNe increases, it will be possible to improve this analysis.

\section{Results}
\label{sec:results}

In this section we examine the luminosities and durations of our model population as given by equations~\eqref{eq:plateau_luminosity} and \eqref{eq:plateau_duration} and make predictions for the observed statistics of LRNe that will be seen by LSST.
We consider three distinct sources of uncertainty in our modelling: uncertainty in the population synthesis prescription; the fraction of the envelope mass that is ejected during a merger; and model uncertainty in the kinetic energy of the ejecta.
The population synthesis variation we consider is whether systems with donors in the Hertzsprung Gap are able to expel their envelope or not, i.e. the optimistic and pessimistic (default) scenario respectively (see section \ref{subsubsec:ce_in_compas}).
For the mass ejected during a merger, we consider $f_\mathrm{m} = 0.05$ (default) and $f_\mathrm{m} = 0.5$.
It is improbable that this quantity is universal for all stellar types and masses, however, no global prescription exists for arbitrary initial masses. Therefore, here we only consider constant values of $f_\mathrm{m}$ to keep our analysis  tractable.
We consider each of the four prescriptions for $E_\mathrm{k}^\infty$ discussed in section~\ref{subsec:LRNe}.
We use as our default model prescription (i) for $E_\mathrm{k}^\infty$ from 
\citep{Ivanova:2013db},
with $\zeta = 10$ for ejections and $\zeta = 1$ for mergers, $f_\mathrm{m} = 0.05$ and pessimistic CE.

\subsection{Population statistics}
\label{subsec:pop_stats}

Here we present summary statistics of our synthetic population of common-envelope events. 
As a sanity check, we compare our results to those of \citet{Politano:2010}, 
who performed a population synthesis study of the merger products from common-envelope evolution  \citep[see also][]{deMink:2013xqa}. 
\citet{Politano:2010} use detailed stellar models for binary evolution, rather than the analytic fits used in COMPAS, taken from \citet{Hurley2000}.  Therefore, 
their stellar structure models allow them to compute the binding energy of the envelope directly, avoiding the need for the $\lambda$ parametrisation described in section~\ref{subsubsec:energy_formalism}.  On the other hand, we are able to consider a broader range of assumptions in our computationally efficient recipe-based population synthesis formalism.  

This comparison is intended only as a guidepost, however, since the approach of 
\citet{Politano:2010} 
differs from ours in several ways:
\begin{itemize}
	\item \citet{Politano:2010}  do not discuss systems that merge while both stars are on the main sequence;
	\item they consider a primary mass range of 
	$0.95 M_\odot \leq M_1 \leq 10 M_\odot$, rather than $1 M_\odot \leq M_1 \leq 100 M_\odot$ used here;
	\item \citet{Politano:2010} use the \citet{1979ApJS...41..513M} IMF, whereas we use the \citet{Kroupa:2000iv} IMF. These IMFs differ for masses $< 1 M_\odot$, but have the same slope above $1 M_\odot$;
	\item \citet{Politano:2010} use a minimum secondary mass of $0.013 M_\odot$, where we use $0.1 M_\odot$ (see section~\ref{subsec:simulation_parameters});
	\item \citet{Politano:2010} use a distribution of orbital periods which is flat in the log of the period
	\citep{Abt:1983}, 
	whereas we use a distribution which is flat in the log of separations out to 1000 AU.  
	They do not state the period of their widest considered binaries.
\end{itemize}
We simulate $5 \times 10^5$ binaries, representing $4.81 \times 10^6 M_\odot$ of star formation (see sections \ref{subsec:simulation_parameters} and \ref{sec:discussion} for the discussion of normalisation). 

52\% of our binaries are wide enough that they never interact and evolve as effectively two single stars, compared to 71\% in \citet{Politano:2010}.
45\% of our simulated binaries undergo some form of unstable mass transfer, compared to 16\% in \citet{Politano:2010}.
6\% of our simulated binaries begin unstable mass transfer when both stars are on the main sequence; COMPAS flags these systems as stellar mergers and does not continue tracking their evolution.
39\% of simulated binaries undergo at least one phase of common-envelope evolution other than main-sequence mergers, with 5\% undergoing two phases of common-envelope evolution.
We now discuss the statistics of the common-envelope phases, excluding those that merge while both stars are on the main sequence.

In our default model, 38\% of common-envelope phases end in stellar mergers, with the remaining 62\% resulting in an ejected envelope, vs.~52\% of common envelopes leading to merger according to \citet{Politano:2010}. 
We find that 76\% of common-envelope phases are initiated by Roche Lobe overflow from the primary onto the secondary, with the remaining 24\% initiated by Roche Lobe overflow from the secondary onto the primary. 
3\% of common envelope phases begin when both stars have developed a core-envelope separation (referred to as a double-core common envelope).

Of the common-envelope interactions that are initiated by the primary, 4\% have donors in the Hertzsprung gap in the \citet{Hurley2000} nomenclature, and the remaining 96\% have donors that are core helium burning, on the giant branch, EAGB, or TPAGB. 
Of those that are initiated by the secondary, 1\% have donors on the main sequence, 6\% have donors in the Hertzsprung gap, and the remaining 92\% have donors that are either core helium burning, on the giant branch, EAGB, or TPAGB.

Common-envelope events involving a compact-object that lead to a merger are likely to appear as very bright supernovae, as the helium core of the donor is disrupted by the compact object and the energy of the outflow is reprocessed by the envelope \citep{Schroder:2019}.   However, we include them as LRN candidates here. 
Such systems form a small subset of our total population ($\approx 2\%$), and do not meaningfully change the characteristics of our predicted plateau distributions.

We repeated the population synthesis simulation with $\alpha =0.5$.
We find that reducing $\alpha$ causes the fraction of common-envelope events leading to merger to increase from 38\% to 51\%.
The stellar properties of the systems that merge or lead to common-envelope ejection do not change appreciably.
Since the majority of detections of LRNe with LSST and ZTF will come from common-envelope ejections, as we show in sections \ref{subsec:plateau}--\ref{subsec:event_rates} below, a lower value of $\alpha$ will reduce our predicted detection rates, but the qualitative features of our predicted distributions do not depend on the value of $\alpha$.

\subsection{Plateau luminosity-duration distributions}
\label{subsec:plateau}

In this subsection, we show predictions for the joint probability distribution function (PDF) of LRN plateau luminosities and durations,
$p(t_\mathrm{p}, L_\mathrm{p})$.
We examine the effect of varying our default model in three ways: turning on the optimistic common envelope assumption; varying the amount of envelope material ejected during a merger; and varying the prescription for the kinetic energy of the ejecta. 

\begin{figure}
\includegraphics[scale=0.55]{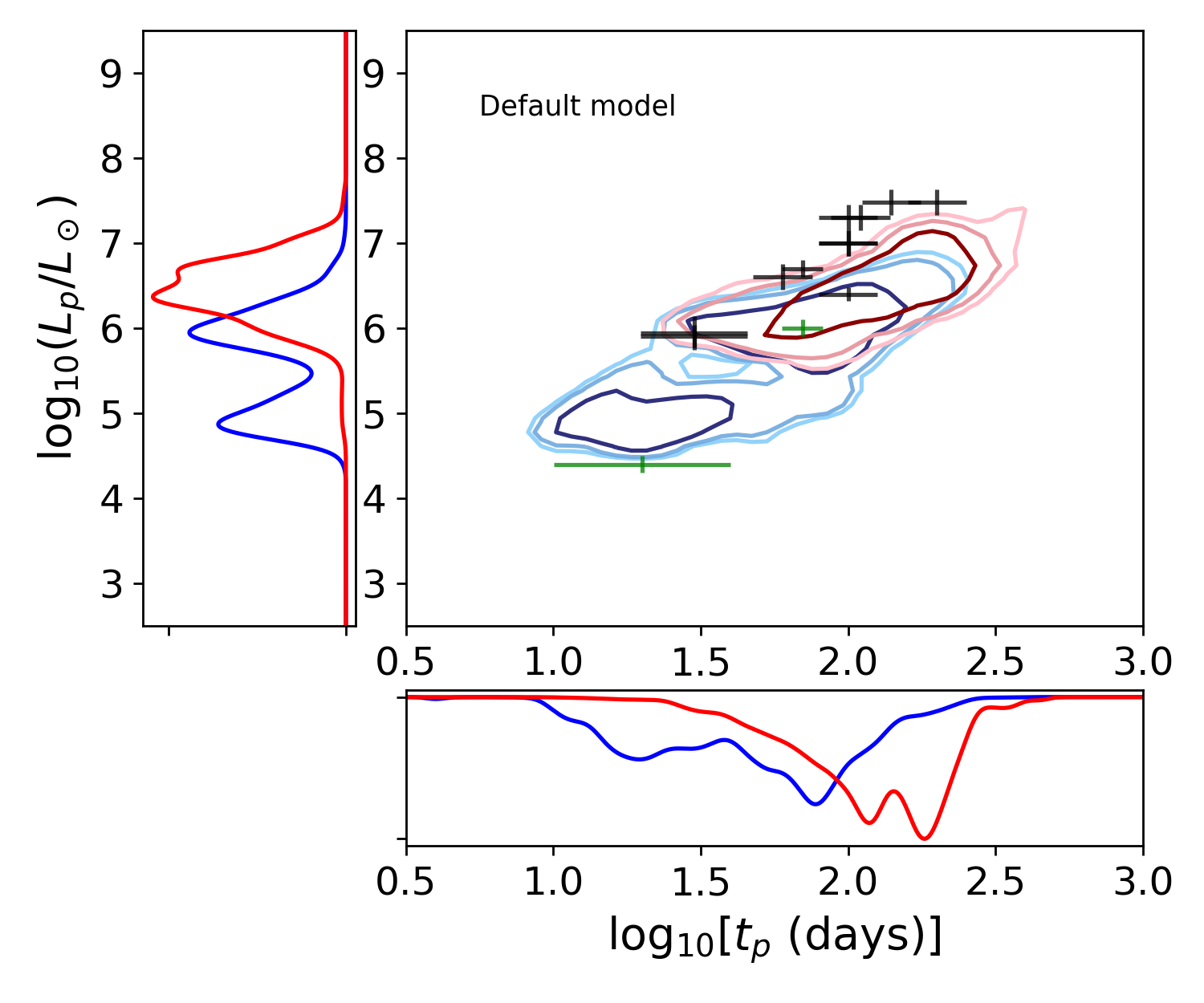}
\caption{
Predicted joint luminosity/duration distribution of luminous red nova plateaux using the default model.
Blue contours show the joint distribution for the intrinsic population in our population synthesis simulation. Red contours show the predicted joint distribution that will be observed by LSST.
Contours enclose 68\%, 90\% and 95\% of the total integrated probability.
Blue and red curves in the margins are the corresponding one-dimensional distributions. 
The crosses (line lengths represent uncertainties) show the observed LRNe which occurred inside (green) and outside (black) the Galaxy.}
\label{fig:default distribution}
\end{figure}

\begin{figure}
\includegraphics[scale=0.55]{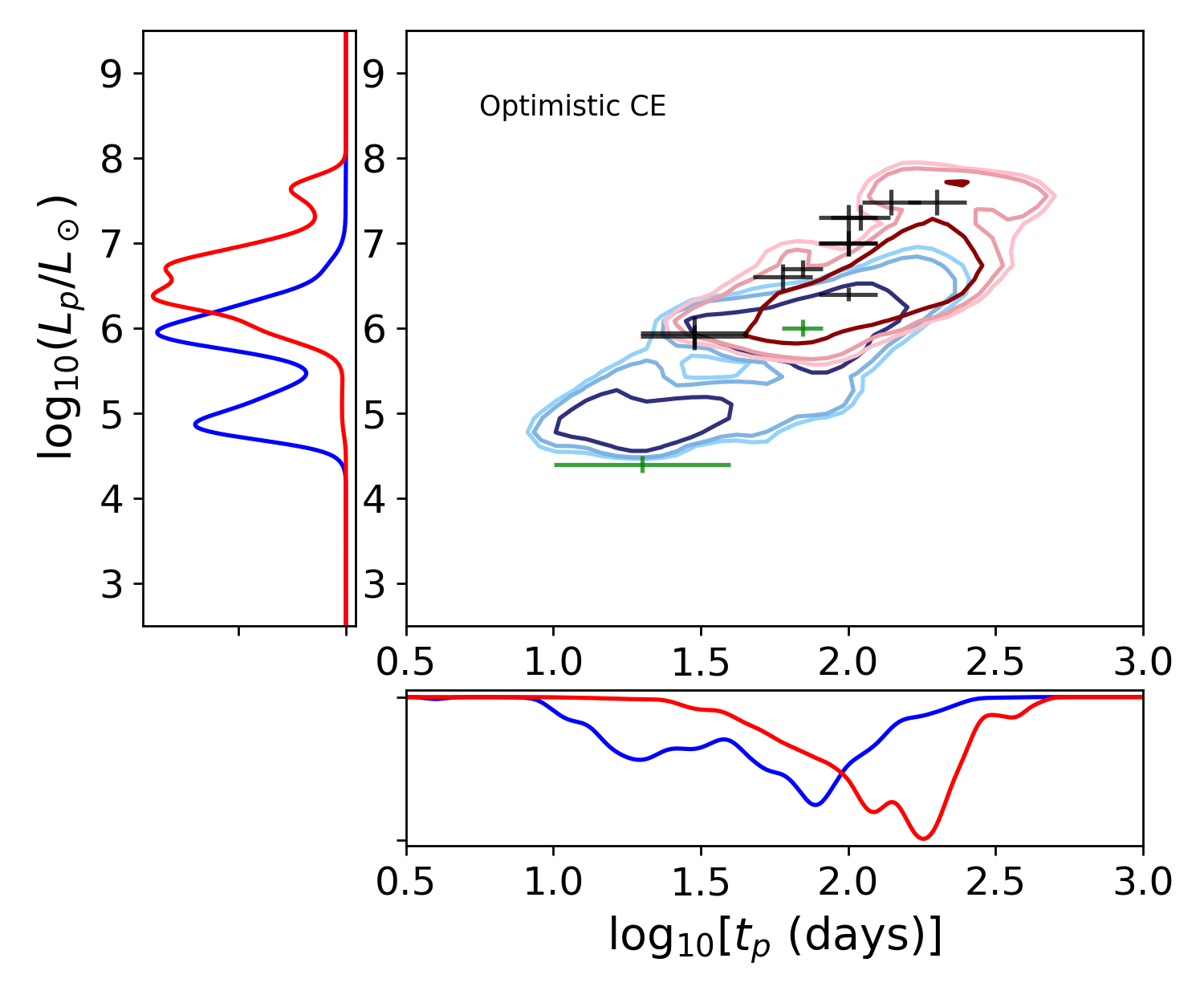}
\caption{Predicted joint luminosity/duration distribution of luminous red nova plateaux.
The contours and symbols have the same meaning as in Figure \ref{fig:default distribution}.
This plot uses the same default ejecta energy model (i) as Figure \ref{fig:default distribution}, but with the optimistic CE assumption.
}
\label{fig:optimistic_CE}
\end{figure}

\begin{figure}
	\includegraphics[scale=0.55]{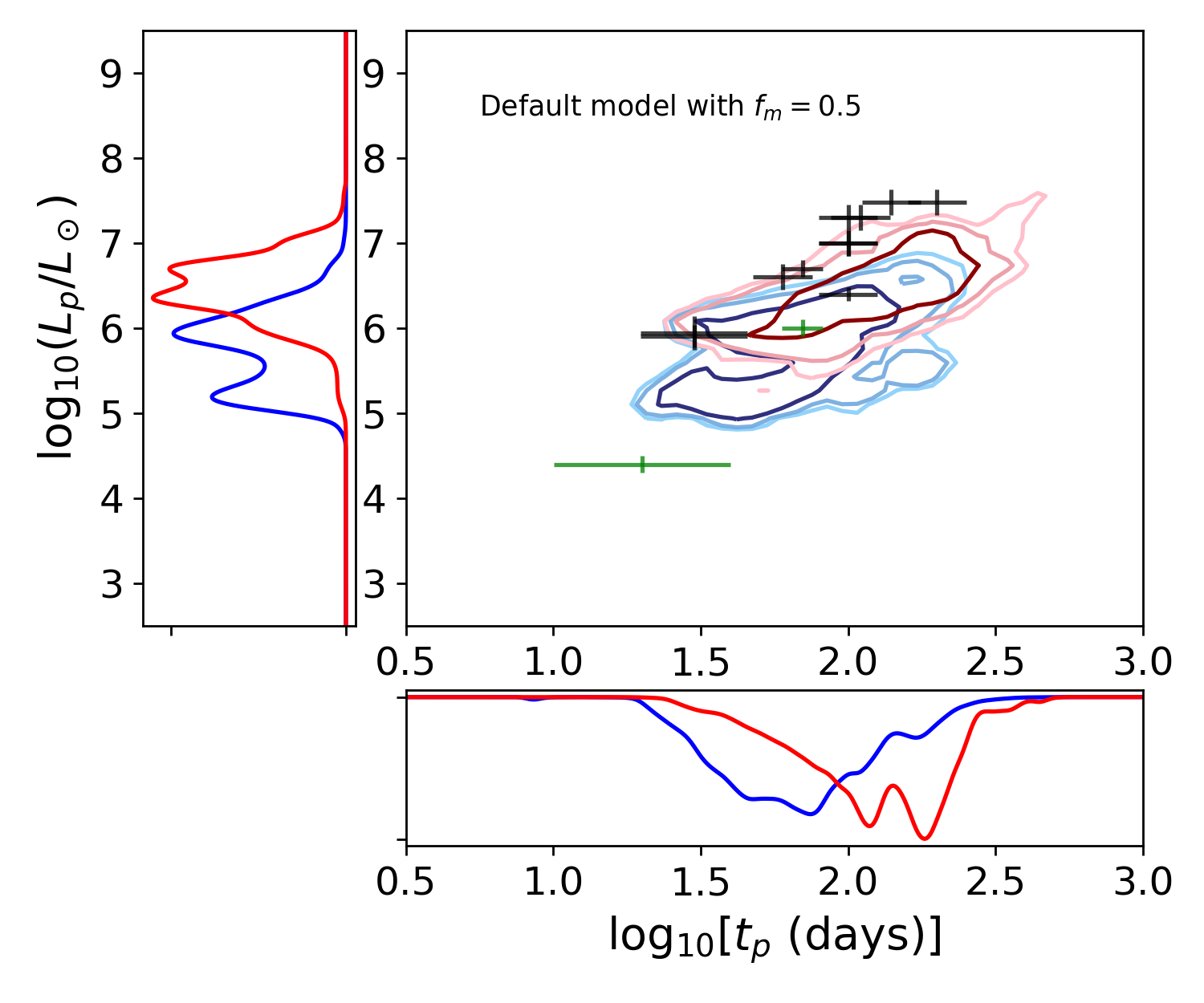} \\
	\caption{Predicted joint luminosity/duration distribution of luminous red nova plateaux.
The contours and symbols have the same meaning as in Figure \ref{fig:default distribution}.
This plot uses the same default ejecta energy model (i) as Figure \ref{fig:default distribution}, but with the merger ejection fraction set to$f_\mathrm{m} = 0.5$.}
	\label{fig:mass_variation}
\end{figure}

\begin{figure}
	\includegraphics[scale=0.55]{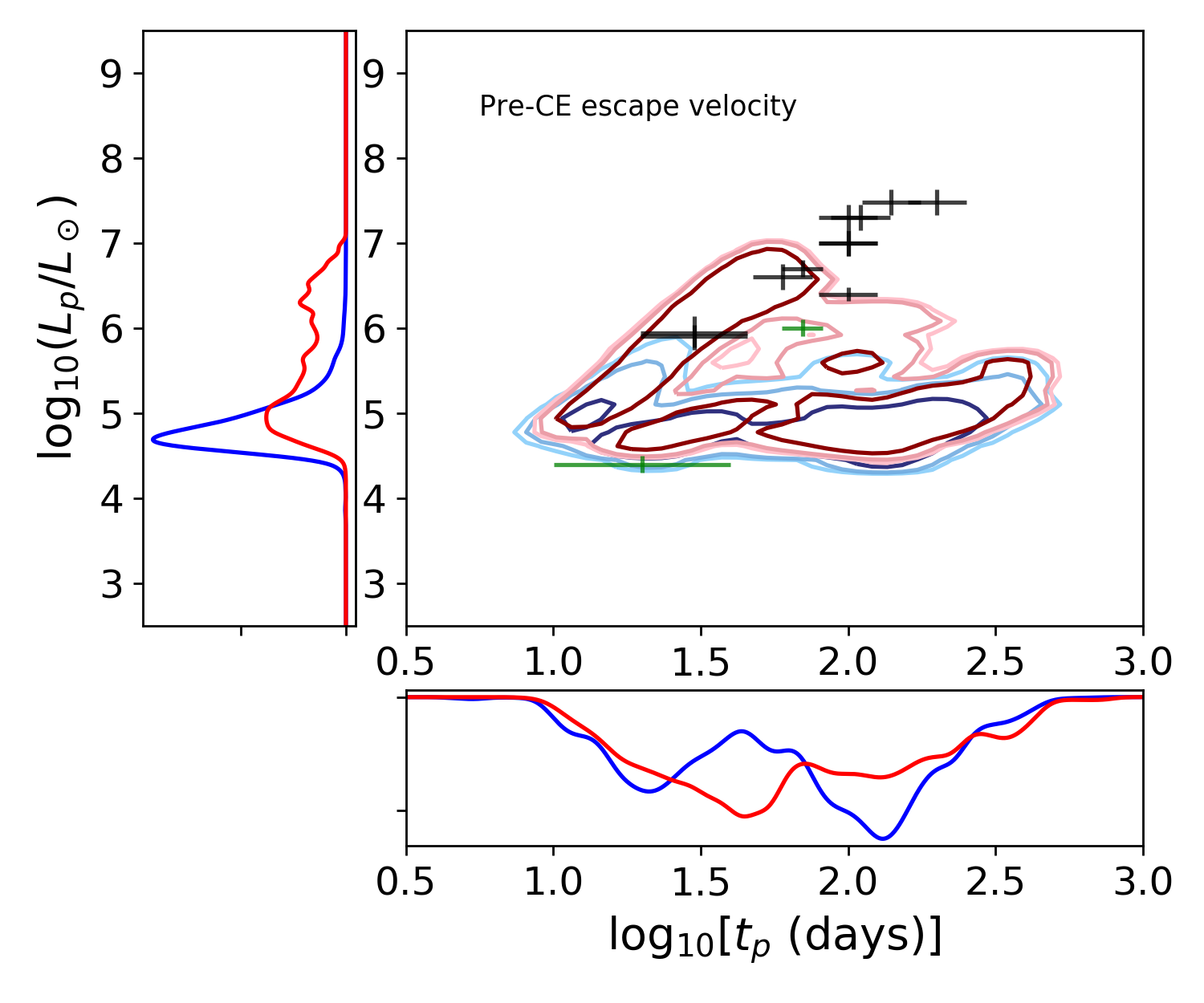} \\
	\includegraphics[scale=0.55]{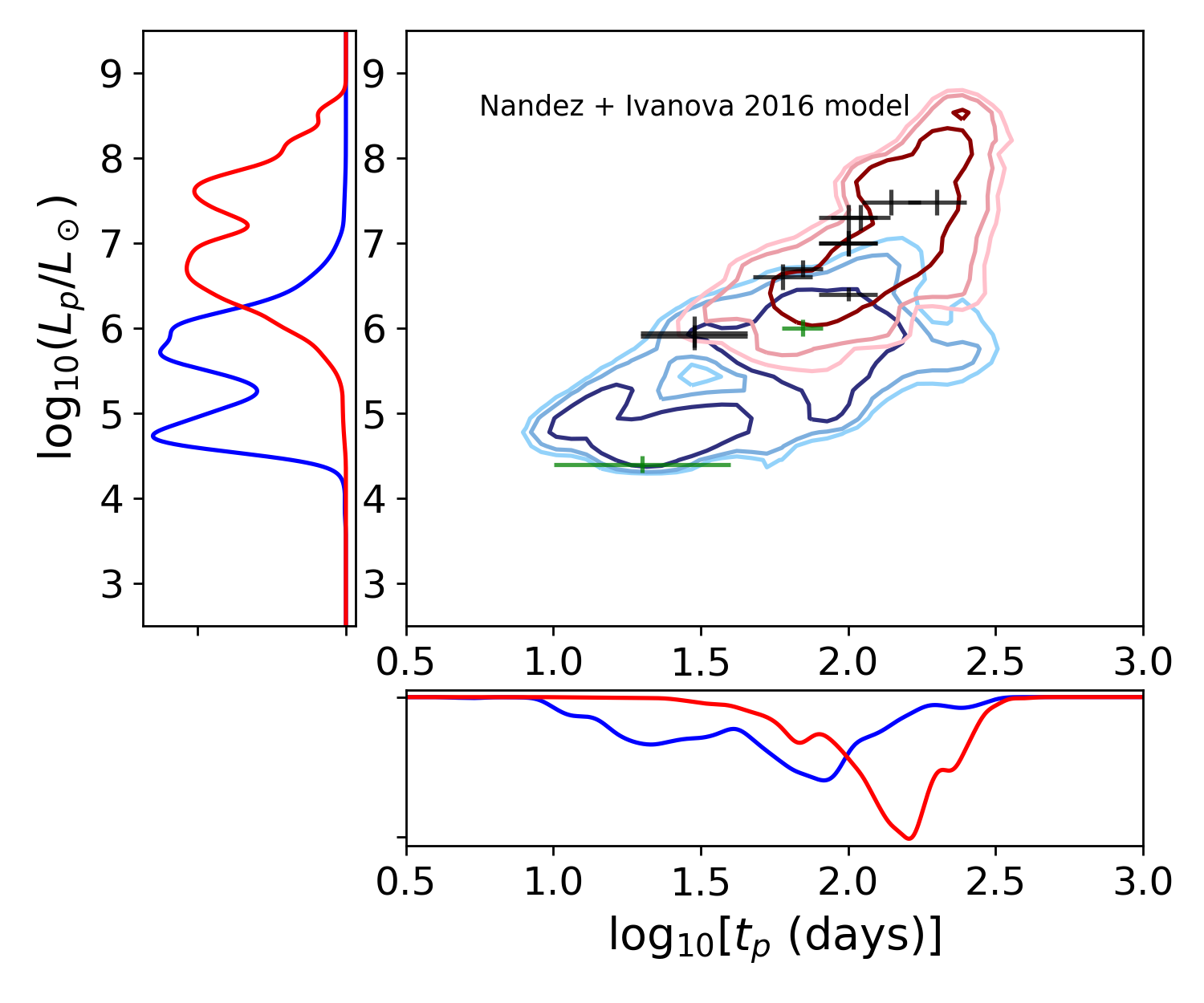} \\
	\includegraphics[scale=0.55]{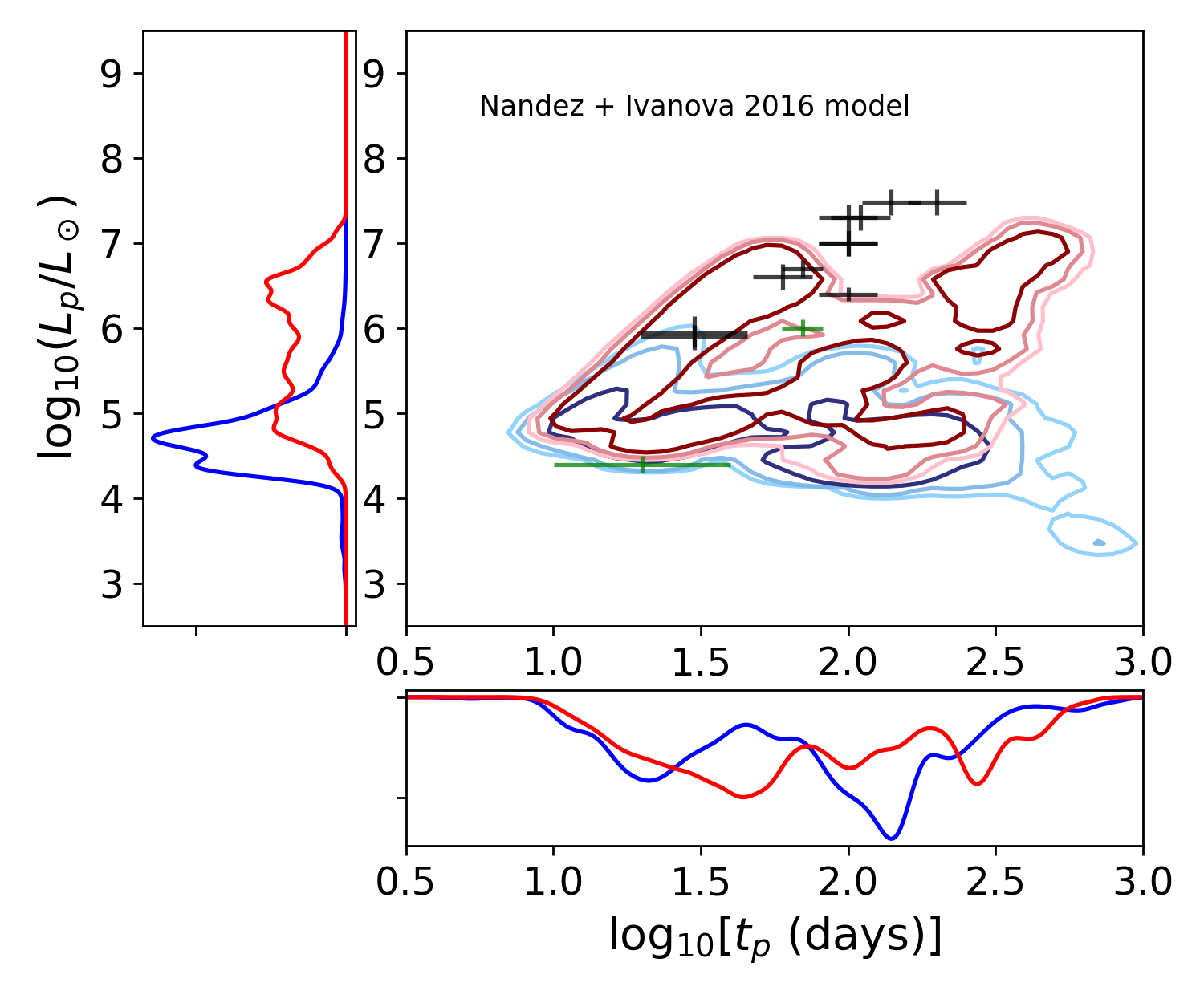}
	\caption{Predicted joint luminosity/duration distributions of luminous red nova plateaux.
The contours and symbols have the same meaning as in Figure \ref{fig:default distribution}.  The panels show predictions for different prescriptions for the ejecta energy $E_\mathrm{k}^\infty$: pre-CE escape velocity prescription (ii-a) in the top panel, post-CE escape velocity prescription (ii-b) in the middle panel, \citet{Nandez2016} prescription (iii) in the bottom panel.}
	\label{fig:ekinf_variations}
\end{figure}

Figure \ref{fig:default distribution} shows the predicted $p(t_\mathrm{p}, L_\mathrm{p})$ for our default model described in section \ref{subsec:LRNe}.
The blue contours show the plateau duration--luminosity distribution of the intrinsic population in our simulation (blue contours), and the red contours show the the selection-biased population, based on the limiting magnitude of LSST (red contours).
Galactic observations, which do not suffer from significant selection effects, are expected to follow the intrinsic population. 
For each distribution, we show contours containing 68\%, 90\%, and 95\% of the total integrated two-dimensional probability density.  
The distributions are smoothed from the Monte Carlo samples with a kernel density estimator: each simulated data point is replaced by a two-dimensional Gaussian `kernel' \citep{Scott1992} in order to produce a smooth estimate of the underlying PDF.
We also show the eight observed LRN plateau luminosities and durations as crosses whose lines represent the uncertainty in each observable.  
Galactic events are in green and extragalactic events in black.
In the margins of the joint distributions we show kernel density estimates of the one-dimensional PDFs $p(t_\mathrm{p})$ and $p(L_\mathrm{p})$ of the intrinsic population (blue curves) and the selection-biased population (red curves).
Figure \ref{fig:default distribution} shows that the intrinsic distribution is bimodal, with the two peaks corresponding to dimmer, shorter-duration LRNe from mergers, and brighter, longer-duration LRNe from common-envelope ejections.
The selection-biased distribution is dominated by the tail of the brightest events from ejections. 
The majority of the LRNe in table \ref{tab:observations} are, within error, inside the outermost contour, however the five brightest events -- NGC 4490--2011OT1, AT 2017jfs, UGC 12307--2013OT1, SNHunt248 and AT 2018hso -- are outside the 95\% contour.
The results for each set of variations are shown in figures \ref{fig:optimistic_CE}, \ref{fig:mass_variation}, and \ref{fig:ekinf_variations}.

The optimistic CE assumption (figure \ref{fig:optimistic_CE}) produces a plateau distribution with brighter events than the pessimistic CE assumption: an additional peak appears at $\log(L_p/L_\odot) \approx 7.5$. 
CE events can be initiated by HG donors when the binary is more compact and the envelope is more tightly bound than for more evolved donors; in our ejecta energy model, this leads to brighter LRNe.  
Only the optimistic CE assumption predicts a distribution which includes the brightest LRNe with the default energy prescription (i).

Figure \ref{fig:mass_variation} shows that the mass fraction ejected during mergers does not substantially change the selection-biased distributions, which are dominated by common-envelope ejections. 
Increasing the mass fraction ejected during mergers does bring the two peaks representing mergers and ejections closer together in the intrinsic luminosity distribution (blue curve), which leads to a poorer match between predictions and the observed Galactic LRNe.
In particular, V1309 Sco lies outside the 95\% integrated probability density contour.

Figure \ref{fig:ekinf_variations} shows that the choice of model for the ejecta energy $E_\mathrm{k}^\infty$ significantly affects our predicted distributions.
The pre-CE escape velocity $E_\mathrm{k}^\infty$ prescription (ii-a) has a unimodal luminosity distribution with reduced ejection energies and yields a significant density near many of the observed events, but does not produce any LRNe as bright as the five brightest.
The reduced ejecta energies in this model deprive the intrinsic distribution of high-luminosity tails, so that the difference between the intrinsic distribution and the luminosity-biased distribution is less pronounced for this model than for other models.
On the other hand, the post-CE escape velocity $E_\mathrm{k}^\infty$ prescription (ii-b) predicts a distribution dominated by brighter, longer duration events than the other models, and has significant density around all observed extragalactic LRNe.  
Meanwhile, the  \citet{Nandez2016} prescription (iii) yields a broad range of plateau durations but generally favours lower luminosities and again does not predict any events as bright as the five brightest observed LRNe. 

\subsection{Event rates}
\label{subsec:event_rates}

\subsubsection{Galactic}
\label{subsubsec:galactic_rates}

Both the intrinsic rate of common-envelope events and the observed rate of LRN transients are highly uncertain. 
The best constraint on the rate of LRNe comes from the number detected within the Galaxy, which, given the long duration and brightness of LRNe, we can assume to be close to a complete sample \citep[though see discussion in][]{Kochanek2014}.
The detection of three Galactic LRNe within the last 25 years gives a Galactic rate of $\sim 0.12$ yr$^{-1}$ per Milky Way equivalent galaxy, or one event per $\sim 17$\,M$_\odot$ of star formation, assuming a Milky Way star formation rate of $2$\,M$_\odot$\,yr$^{-1}$
\citep{Licquia2015a}.  
\citet{Ofek:2007nw} 
inferred a 95\%-confidence lower limit of 0.019 yr$^{-1}$ on the rate based on the observations of V838 Mon and V4332 Sgr. 
These values match the rate from our simulation of one event per $19.1 M_\odot$ of star formation, or 0.1 yr$^{-1}$ in the Milky Way.  
They also roughly agree with a previous population study by 
\citet{Kochanek2014}, who modelled the Galactic rate of common-envelope events to be 0.2 yr$^{-1}$.

\begin{figure}
	\includegraphics[width=\columnwidth]{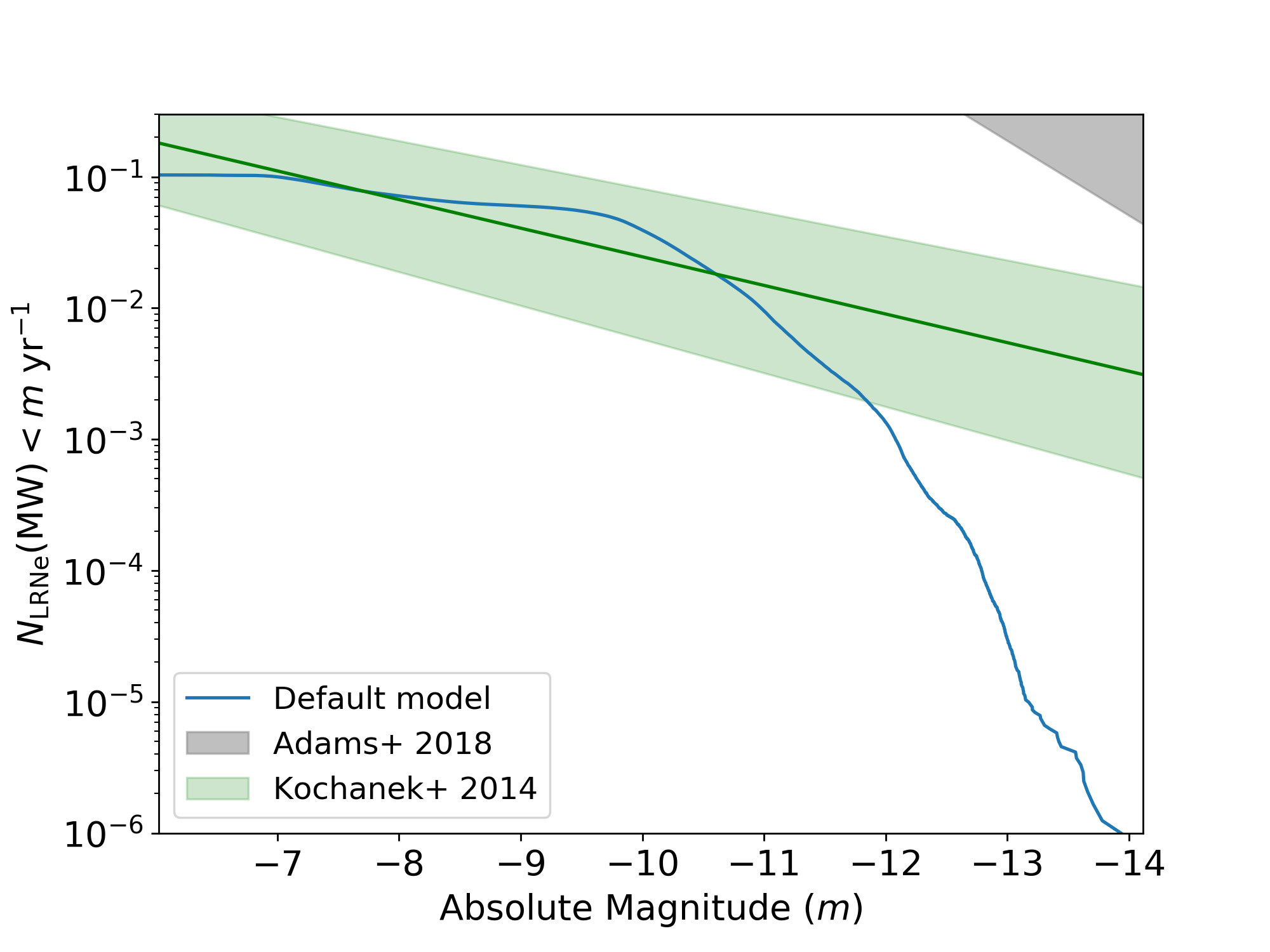}
	\caption{Rate of Galactic LRNe brighter than a given absolute magnitude $m$ as predicted by our default model.  The gray shaded region corresponds to the observational upper limit from \citet{Adams:2018PASP}, and the green line and shaded region correspond to empirical constraints from Galactic LRNe in \citet{Kochanek2014}.
	Note that the comparison is not exact, as our magnitudes are bolometric while those of \citet{Adams:2018PASP} and \citet{Kochanek2014} are in the $I$ band.}
	\label{fig:galactic_rate}
\end{figure}

In Figure~\ref{fig:galactic_rate} we show the intrinsic (i.e. Galactic) rate of LRNe as a function of absolute magnitude $(m)$ predicted by our default model\footnote{We use $m$ for the absolute magnitude to avoid confusion with mass $M$.}.
We also show the observational upper limit constraints from \cite{Adams:2018PASP} as the grey shaded region, and the empirical constraints from \cite{Kochanek2014} as the green shaded region.
Our model agrees with both constraints, except for the brightest events with $M \lesssim -12$, where it diverges from the rate of \cite{Kochanek2014}.
However, their rate is derived from only three Galactic observations, and is extrapolated for bright events. 

\subsubsection{Extragalactic}
\label{subsubsec:extragalactic_rates}

Outside the Galaxy, assuming a star formation rate of $0.015$\,$M_\odot$\,Mpc$^{-3}$\,yr$^{-1}$ 
\citep{Madau2014}, our predicted local average LRN rate is $\sim 8 \times 10^{-4}$\,Mpc$^{-3}$\,yr$^{-1}$. 
Comparing our volumetric LRN rate with the rate of formation of \acp{BBH} which merge within a Hubble time, 24--112 Gpc$^{-3}$ yr$^{-1}$
\citep{gwtc},
we predict that LRNe are $\sim 10^4$ times more common than the formation of merging \acp{BBH}, and hence, even if the majority of \acp{BBH} are formed through common-envelope evolution, their progenitors are unlikely to form an appreciable sub-population of LRNe.

\subsubsection{Upcoming surveys - LSST and ZTF}
\label{subsubsec:upcoming_surveys}

We make predictions for the populations of LRNe observable in future surveys with LSST \citep{LSSTScience:2009} and the Zwicky Transient Facility (ZTF) \citep{Graham:2019PASP,Bellm:2019PASP}.

\begin{figure}
\includegraphics[width=\columnwidth]{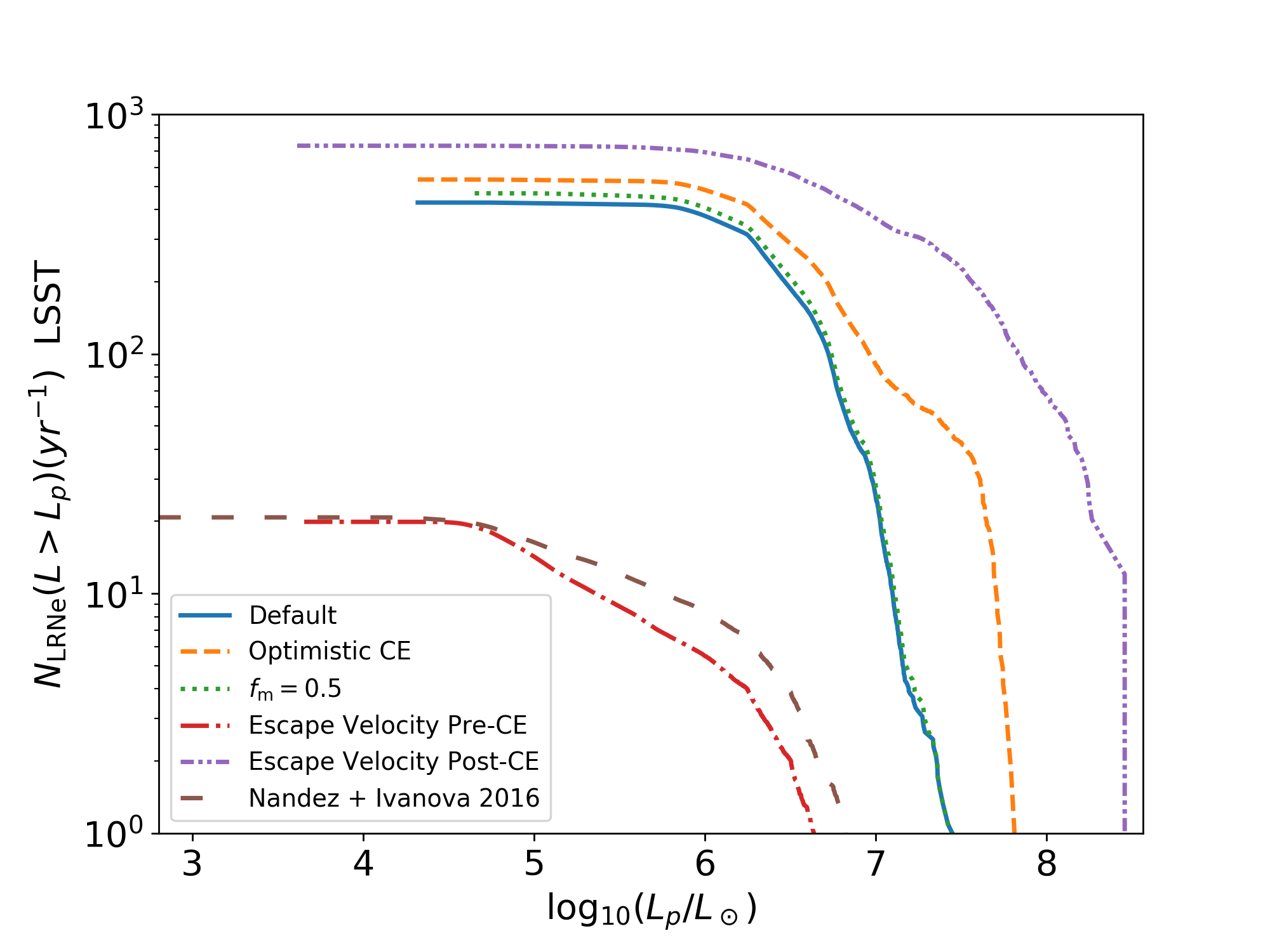}
\includegraphics[width=\columnwidth]{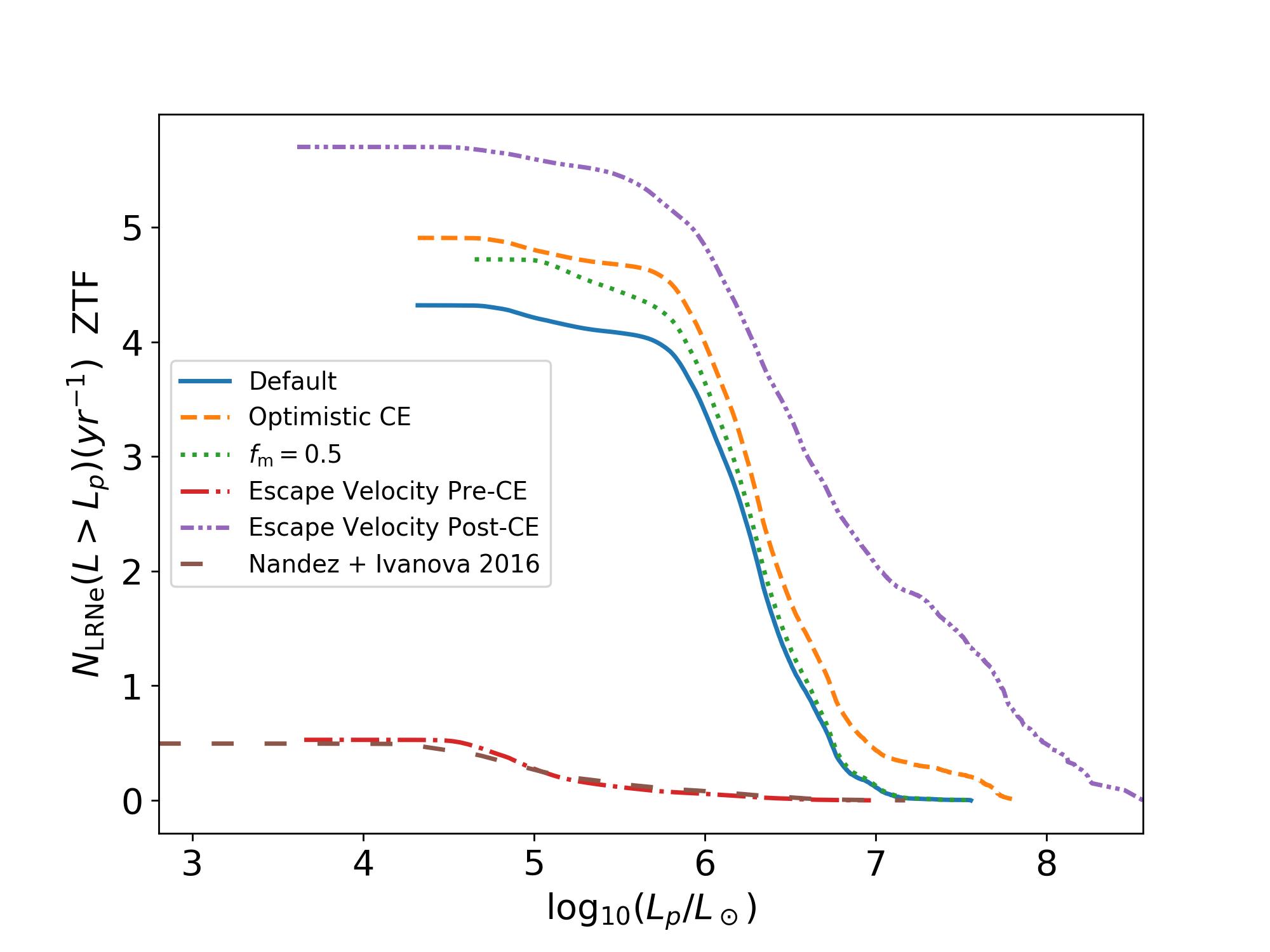}
\caption{Top panel: cumulative rate of LRNe observed by LSST brighter than a given plateau luminosity $L_p$ as predicted by our default model and model variations described in section \ref{subsec:LRNe}. 
Bottom panel: cumulative rate of LRNe observed by ZTF (while the scaling between curves corresponding to different models is the same as in the top panel, but vertical axis is linear rather than logarithmic for clarity) .
}
\label{fig:event_rates}
\end{figure}

In the top panel of Figure~\ref{fig:event_rates} we show the predicted detection rate for LRNe with LSST as a function of plateau luminosity for the model variants described in section \ref{subsec:LRNe}. 
The bottom panel of Figure~\ref{fig:event_rates} shows the same for ZTF.
The default model predicts 430 LRN detections per year with LSST, and we summarise the rates from our model variations in Table
\ref{tab:models}.
\begin{table}
	\centering
	\begin{tabular}{@{}lll@{}}
		\toprule
		Model Name  & LSST Rate & ZTF Rate \\ 
		&  [yr$^{-1}$] &  [yr$^{-1}$]  \\
		\midrule
		Default            &          430   & 4.3  \\
		Optimistic CE   &      530 & 4.9 \\
		$f_\mathrm{m} = 0.5$    &      470  & 4.7 \\
		(ii-a) Pre-CE escape velocity   &      20 & 0.5  \\
		(ii-b) Post-CE escape velocity  &      740 & 5.7  \\
		(iii)  Nandez \& Ivanova 2016  &     20 & 0.5   \\
		\bottomrule
	\end{tabular}
	\caption{Predicted LRN detection rates with LSST and ZTF for the models we consider in this work.
		\label{tab:models}}
\end{table}
Apart from the pre-CE escape velocity model (ii-a) and the \citet{Nandez2016} model (iii(), all our LSST detection rate estimates fall within the previously predicted LSST detection rate envelope of 80--3400 yr$^{-1}$ \citep{LSSTScience:2009}.
With our default model, we predict that ZTF will detect \textbf{$\approx 4$} LRN each year, consistent with \cite{Adams:2018PASP}, and between our various models we predict between \textbf{$\approx 0.5$--$6$} LRNe per year with ZTF.

We find that the brightest LRNe have luminosities $6.5 < \log_{10}(L_p / L_\odot)< 8.5$ (absolute bolometric magnitudes $-11.5>m>-16.5$), depending on our model assumptions.
The brightest observed LRN NGC4490-OT \citep{Smith:2016qtr,Pastorello2019} had an absolute \textbf{bolometric} magnitude of $\sim -14$, suggesting it was among the brightest such events we can expect to observe.

\section{Discussion}
\label{sec:discussion}

We have considered a range of models for connecting CE events to LRNe, although this is not an exhaustive list of plausible variations.  
The current sample of observed Galactic and extragalactic LRNe is already constraining. 
Only our default ejecta energy model (i) with optimistic CE and the post-CE escape velocity ejecta energy model (ii-b) can plausibly explain all events. 
The rest of our models fail to predict LRNe as bright as NGC 4490--2011OT1, AT 2017jfs, UGC 12307--2013OT1, SNHunt248 and AT 2018hso; and, in the case of the default model with $f_\mathrm{m} = 0.5$, fail to predict LRNe as dim as V1309 Sco.
Of course, we only consider a limited set of plausible variations for LRN energetics and common-envelope physics. 
Our goal is not to tweak the models in order to match the observations (which is possible, but uninformative), but rather to explore the range of predicted detection rates and the science that can be done with a growing data set.

Our plausible models predict LRN detection rates with LSST of roughly 500 detections per year, in line with previous predictions, and LRN detection rates of roughly 5 per year with ZTF. 
We also predict a pronounced difference between the intrinsic population of LRNe plateaux and the observed population.  
The intrinsic population has roughly similar amounts of LRNe from mergers and common-envelope ejections, which results in a bimodal plateau luminosity distribution.
The extragalactic selection-biased population, however, is dominated by the brightest events, which are almost exclusively due to envelope ejections. 

In determining the observed rate of LRNe, we ignored the time delay between star formation and the common envelope event.   Since our model for the Milky Way assumes a constant star formation rate and metallicity, this time delay does not affect our predictions for the population of LRNe observable in the Galaxy.  
\begin{figure}
\includegraphics[width=\columnwidth]{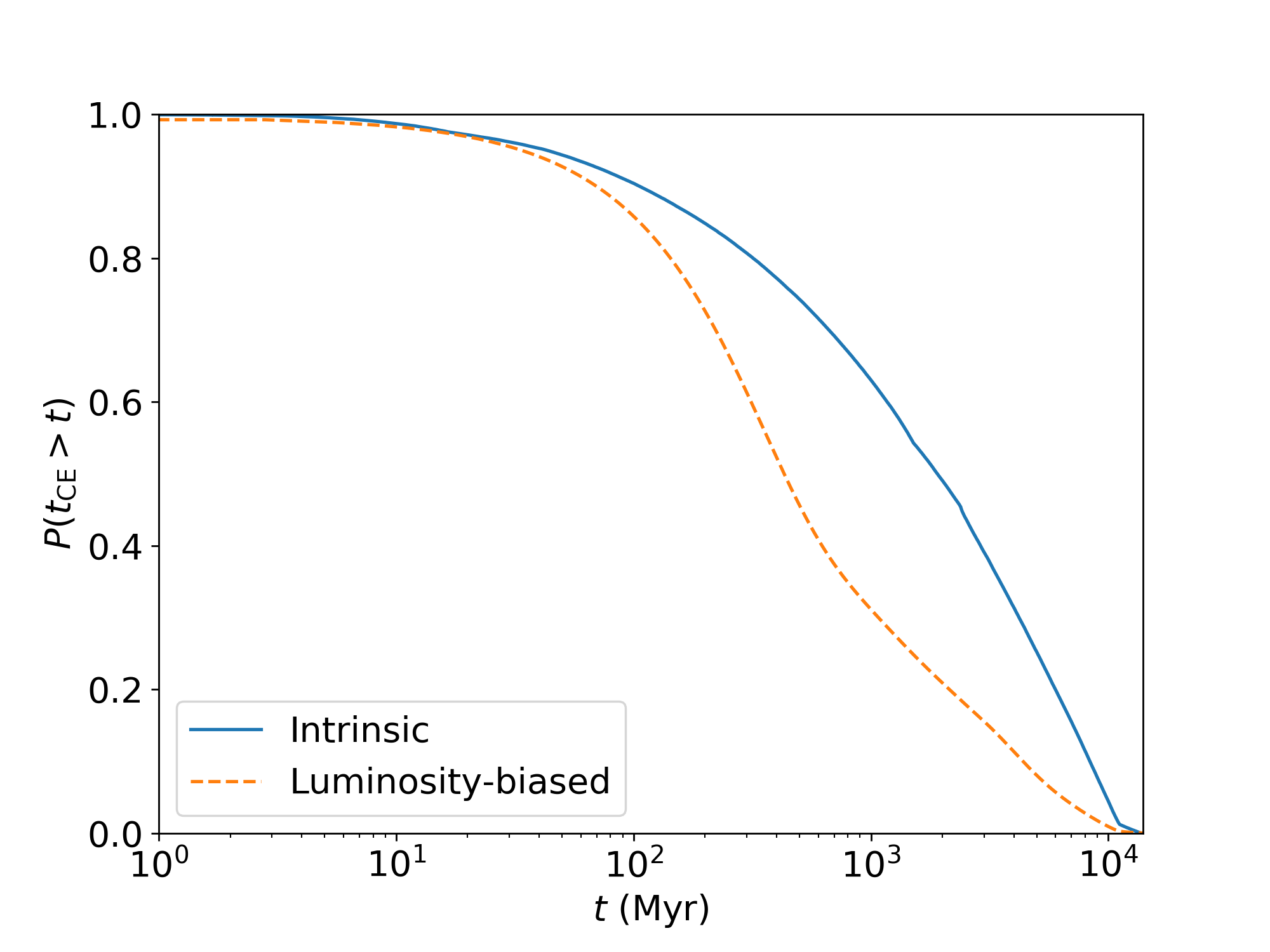}
\caption{Cumulative distribution of the delay time between star formation and CE event. The blue solid curve shows the intrinsic distribution, while the orange dashed curve includes the observational bias.}
\label{fig:time_to_ce_cdf}
\end{figure}

\begin{figure}
\includegraphics[width=\columnwidth]{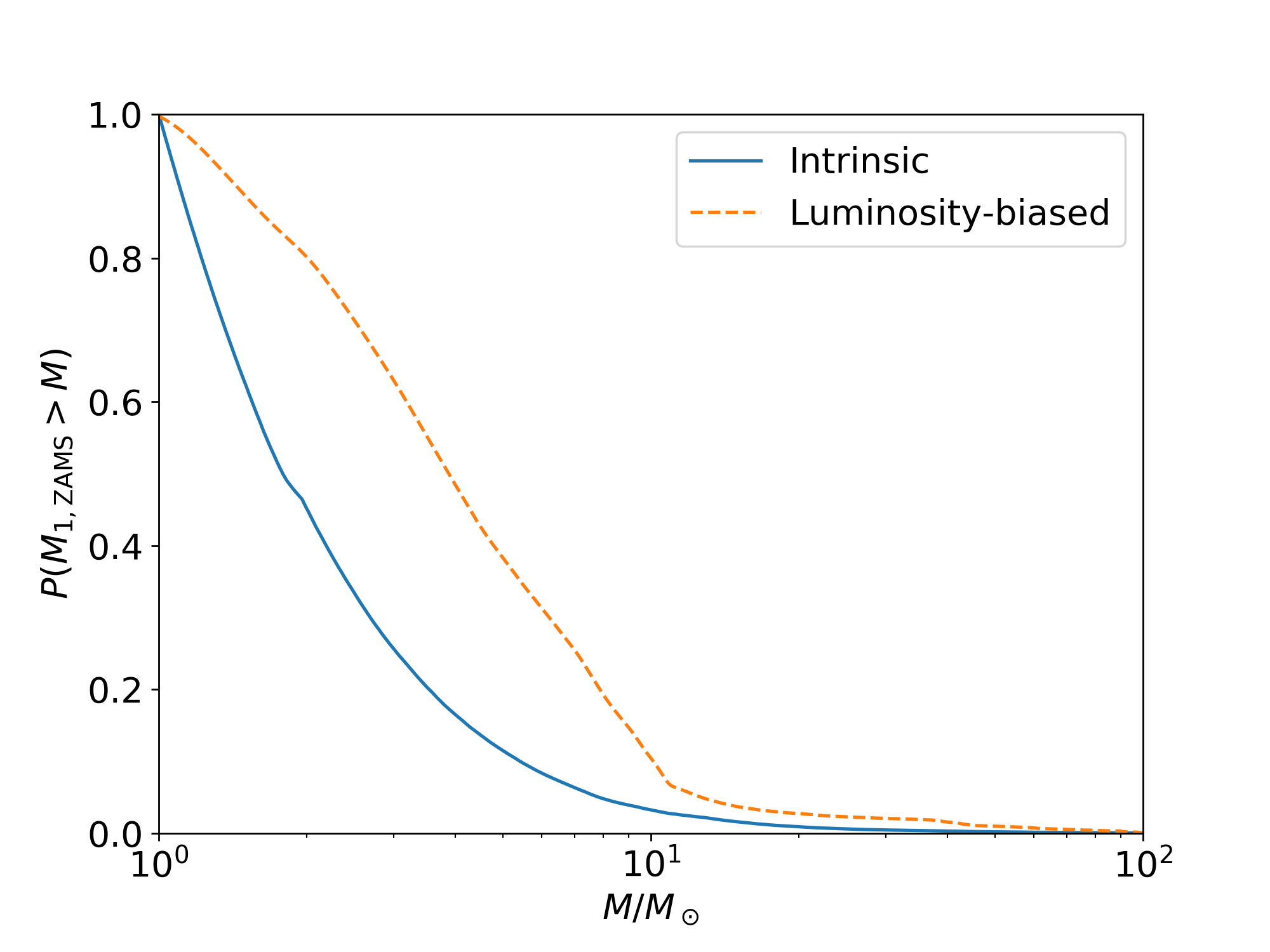}
\caption{Cumulative distribution of the birth mass of the initially more massive star $m_1$ in our default model. 
The blue solid curve shows the intrinsic distribution, while the orange dashed curve includes the selection bias.}
\label{fig:primary_mass_cdf}
\end{figure}

On the other hand, for extragalactic LRNe, a binary undergoing a common envelope event observable today could have formed up to $\sim 10$\,Gyr ago (see Figure~\ref{fig:time_to_ce_cdf}), corresponding to a redshift of $z \approx 2$, when the global star formation rate was a factor $\sim 10$ higher \citep{Madau2014}.  
However, this long tail of the delay-time distribution is due to low-mass stars, which typically produce dim LRNe.  This can be understood analytically as follows.  
The masses of stars are distributed according to the IMF, which for the masses we consider, $M > 1$\,M$_\odot$, scales as $M^{-2.3}$ \citep{Salpeter:1955ApJ,Kroupa:2000iv}, with low mass stars being the most common (see Figure~\ref{fig:primary_mass_cdf}). The lifetime of a low mass star scales with its mass as $t \sim M^{-2.5}$. The distribution of main sequence lifetimes thus scales as $\mathrm{d}N/\mathrm{d}t \sim t^{-0.5}$ with a median delay time of $\sim 2$\,Gyr.  Moreover, the brightest, luminosity-selected events are dominated by  more massive, rapidly evolving stars  (see Figure~\ref{fig:primary_mass_cdf}).  Consequently, the luminosity-selected events have median delay times of well under a Gyr, as shown in Figure~\ref{fig:time_to_ce_cdf}.  Since the cosmic star formation rate did not change significantly on this timescale \citep{Madau2014}, we can use the simplifying assumption of a constant star formation rate throughout.   On the other hand, variations in the metallicity of star-forming gas \citep{Madau2014,Neijssel:2019,Chruslinska:2019} are potentially important, and have not been addressed here.

For simplicity, we have assumed that all stars form in binaries in our models, although only around $\sim 50\%$ of these binaries interact (exchange mass) at some point during their lives. 
Observations have shown that stellar multiplicity appears to be a function of stellar mass. 
The majority of massive stars are born in binaries \citep[e.g.][]{Sana:2012,Moe:2017ApJS}, but the fraction of stars with companions decreases for lower-mass stars \citep{Raghavan:2010ApJS}.
Since the majority of luminosity-selected LRNe arise from binaries with $M_\mathrm{1\,ZAMS} > 4\, M_\odot$ stars (see Figure~\ref{fig:primary_mass_cdf}), neglecting this trend introduces an uncertainty in quoted event rates that is well below other sources of modelling uncertainty. 

Another caveat is the possibility that some of the bright, long-duration events predicted here will be self-obscured by the optically thick ejecta, and will be re-processed into the infrared \citep[c.f.][]{Minniti2017,Kasliwal:2017fqr,Oskinova:2018A&AL}.  Of course, the existence of very bright LRNe such as NGC 4490-OT2011 \citep{Smith:2016qtr} would argue that at least some LRNe avoid this fate.  However, NGC 4490-OT2011 appears to resemble SN 2008S \citep{2008CBET.1234....1A}, which is not a consensus LRN candidate.

\citet{Clayton:2017} argued that some common envelope events may have multiple ejections.  This would increase the rates relative to those we quote, but could decrease the luminosity and duration of individual LRNe, especially if later ejections are obscured by the material emitted during earlier ones.


Massive red supergiants (RSGs) in low-metallicity environments have been invoked to explain the formation through common-envelope evolution \citep[e.g.][]{Belczynski:2016obo,Stevenson:2017tfq,Stevenson:2019} of the merging binary black holes being observed by Advanced LIGO and Virgo \citep{gwtc}\footnote{The CE event in the illustrative example of the plausible formation of the source of the first gravitational-wave detection GW150914 by \citet{Belczynski:2016obo} would yield an LRN with a plateau luminosity of $\sim 2.5 \times 10^7 L_\odot$ in our default model. However, as we discuss in Section~\ref{subsubsec:extragalactic_rates}, LRNe associated with \ac{BBH} formation are sufficiently rare that we do not expect them to form an observable subpopulation.}. However, there are no massive RSGs observed in the Milky Way and the Magellanic Clouds with a luminosity greater than $\sim 10^{5.5}$\,L$_\odot$ \citep{1979ApJ...232..409H,Levesque:2017}. Stellar models predict that the most massive RSGs correspond to single stars with initial masses of $\sim 40$\,M$_\odot$ \citep[e.g.][]{2012A&A...537A.146E,Sanyal:2017,Groh:2019}. In our model, stars with initial masses $\gtrsim 40$\,M$_\odot$ have high mass loss rates and do not form RSGs at solar metallicity. 

Using our model, we can estimate what the luminosity of a LRN with a massive RSG donor would look like.  
Taking approximate sample parameters of a $40 M_\odot$, $10^{5.4} L_\odot$ donor with a temperature of $4000$ K and hence a radius of $\sim 1000 R_\odot$ which ejects a 20 $M_\odot$ envelope during a CE event would yield a $1.3 \times 10^7 L_\odot$ LRN according to the default ejecta energy prescription (i).  
This lies at the upper end of our predictions using this model, and suggests that a population of more luminous events with red progenitors would point to the existence of massive RSGs. RSGs can have higher masses and smaller radii for a given mass at lower metallicity.  
The plateau luminosity scales as donor radius to the $-1/6$ power in model (i) if all other properties remain unchanged, and as mass to the $7/6$ power, so LRNe with massive RSG donors would be brighter in low-metallicity environments. 
However, alternative models, such as the post-CE energy ejecta model (ii-b), predict events of this luminosity from COMPAS populations that do not have massive RSGs, consistent with the \textbf{$3 \times 10^7 L_\odot$} LRN NGC 4490-OT2011 that is associated with a blue progenitor.

Observations of the significant population of LRNe that will be accessible with LSST will provide insights on CE physics and the evolution of massive stars that will be complementary to existing observations.  Given the critical importance of CE events to massive binary evolution, LRNe may play an important role in elucidating massive binaries as progenitors of gravitational-wave sources.

\section*{Acknowledgments}
GH visited the University of Birmingham on an Alan Kenneth Head Travelling Scholarship from the University of Melbourne.
AVG  acknowledges funding support from CONACYT.
GH and SS are supported  by  the  Australian Research Council Centre of Excellence for Gravitational  Wave  Discovery  (OzGrav), through  project number CE17010000.  SJ is grateful for partial support of this work via a grant from the Simons Foundation.
NI, SJ, and IM acknowledge that a part of this work was performed at the KITP, which is supported in part by the National Science Foundation under Grant No. NSF PHY-1748958. NI, SJ, and IM acknowledge that a part of this work was performed at the Aspen Center for Physics, which is supported by National Science Foundation grant PHY-1607611. 
IM is an Australian Research Council Future Fellow, project FT190100574.
NI acknowledges support from CRC program and funding from NSERC Discovery. 

\bibliographystyle{mnras}
\bibliography{lrn_bib}

\label{lastpage}

\end{document}

%% file: observations_table.tex
\begin{table*}
\centering
\begin{tabular}{lllp{8cm}}
\toprule
Event Name      & Duration {[}days{]} & $L/L_\odot$ \\ \midrule
V838 Mon (G)       &           $70 \pm 10$        &        $(1 \pm 0.2) \times 10^6$      & \citet{Tylenda:2005}     \\
V1309 Sco  (G)    &       $20 \pm 10$       &   $(2.5 \pm 0.5) \times 10^{4}$  & \citet{Tylenda:2011}    \\
M85-OT (E)          &      $70 \pm 10$       &    $(5 \pm 1) \times 10^6$       & \citet{Kulkarni:2007kf,Pastorello:2007em,Rau:2006vc,Ofek:2007nw}\\
NGC 4490$-$2011OT1 (E) &     $200 \pm 40$       &      $(3 \pm 1) \times 10^7$        & \citet{Smith:2016qtr,Pastorello2019}      \\
M31 2015 (E)       &        $30\pm 10$          &   $8.7^{+3.3}_{-2.2} \times 10^5$         & \citet{Kurtenkov:2015,Williams:2015,Macleod2017} \\
M31 RV 1988 (E)  &           $30 \pm 10$         &      $(8 \pm 2) \times 10^5$      & \citet{Mould1990}     \\
M101 OT2015 (E)   &          $100 \pm 20$           &        $(2.5 \pm 1) \times 10^6$        & \citet{Goranski2016,Blagorodnova:2017,Lipunov2017,Pastorello2019}     \\
NGC 3437$-$2011OT1 (E)  &  $100 \pm 20$  &  $(1 \pm 0.3) \times 10^7$  &  \citet{Pastorello2019}  \\
UGC 12307$-$2013OT1 (E)  &  $50 \pm 20$  &  $(2 \pm 0.5) \times 10^7$  &  \citet{Pastorello2019}  \\
SNhunt248 (E)  &  $110 \pm 20$   &  $(2 \pm 1) \times 10^7$  &  \citet{Kankare2015,Mauerhan2018,Pastorello2019}  \\
AT 2017jfs (E)  &  $140 \pm 20$  &  $(3 \pm 1) \times 10^7$  &  \citet{Pastorello2019}  \\
AT 2018hso (E)  &  $100 \pm 20$  &  $(1 \pm 0.3) \times 10^7$  &  \citet{Cai2019,Pastorello2019}  \\
SN 1997bs (E)  &  $60 \pm 20$  &  $(4 \pm 2) \times 10^6$   &  \citet{VanDyk2000,Pastorello2019}  \\
\bottomrule
\end{tabular}
\caption{Summary of observed LRNe with reliably inferred absolute plateau luminosities and durations. 
Events labelled (G) occurred within the Galaxy and events labelled (E) were of extragalactic origin.
\label{tab:observations}}
\end{table*}

%% file: luminous_red_novae.bbl
\begin{thebibliography}{}
\makeatletter
\relax
\def\mn@urlcharsother{\let\do\@makeother \do\$\do\&\do\#\do\^\do\_\do\%\do\~}
\def\mn@doi{\begingroup\mn@urlcharsother \@ifnextchar [ {\mn@doi@}
  {\mn@doi@[]}}
\def\mn@doi@[#1]#2{\def\@tempa{#1}\ifx\@tempa\@empty \href
  {http://dx.doi.org/#2} {doi:#2}\else \href {http://dx.doi.org/#2} {#1}\fi
  \endgroup}
\def\mn@eprint#1#2{\mn@eprint@#1:#2::\@nil}
\def\mn@eprint@arXiv#1{\href {http://arxiv.org/abs/#1} {{\tt arXiv:#1}}}
\def\mn@eprint@dblp#1{\href {http://dblp.uni-trier.de/rec/bibtex/#1.xml}
  {dblp:#1}}
\def\mn@eprint@#1:#2:#3:#4\@nil{\def\@tempa {#1}\def\@tempb {#2}\def\@tempc
  {#3}\ifx \@tempc \@empty \let \@tempc \@tempb \let \@tempb \@tempa \fi \ifx
  \@tempb \@empty \def\@tempb {arXiv}\fi \@ifundefined
  {mn@eprint@\@tempb}{\@tempb:\@tempc}{\expandafter \expandafter \csname
  mn@eprint@\@tempb\endcsname \expandafter{\@tempc}}}

\bibitem[\protect\citeauthoryear{{Abbott} et~al.,}{{Abbott}
  et~al.}{2019}]{gwtc}
{Abbott} B.~P.,  et~al., 2019, \mn@doi [Physical Review X]
  {10.1103/PhysRevX.9.031040}, \href
  {https://ui.adsabs.harvard.edu/abs/2019PhRvX...9c1040A} {9, 031040}

\bibitem[\protect\citeauthoryear{{Abt}}{{Abt}}{1983}]{Abt:1983}
{Abt} H.~A.,  1983, \mn@doi [\araa] {10.1146/annurev.aa.21.090183.002015},
  \href {http://adsabs.harvard.edu/abs/1983ARA%26A..21..343A} {21, 343}

\bibitem[\protect\citeauthoryear{Adams, Kochanek, Prieto, Dai, Shappee  \&
  Stanek}{Adams et~al.}{2016}]{Adams:2015xjy}
Adams S.~M.,  Kochanek C.~S.,  Prieto J.~L.,  Dai X.,  Shappee B.~J.,   Stanek
  K.~Z.,  2016, \mn@doi [Mon. Not. Roy. Astron. Soc.] {10.1093/mnras/stw1059},
  460, 1645

\bibitem[\protect\citeauthoryear{{Adams} et~al.,}{{Adams}
  et~al.}{2018}]{Adams:2018PASP}
{Adams} S.~M.,  et~al., 2018, \mn@doi [\pasp] {10.1088/1538-3873/aaa356}, \href
  {https://ui.adsabs.harvard.edu/abs/2018PASP..130c4202A} {130, 034202}

\bibitem[\protect\citeauthoryear{{Arbour} \& {Boles}}{{Arbour} \&
  {Boles}}{2008}]{2008CBET.1234....1A}
{Arbour} R.,  {Boles} T.,  2008, Central Bureau Electronic Telegrams, \href
  {http://adsabs.harvard.edu/abs/2008CBET.1234....1A} {1234}

\bibitem[\protect\citeauthoryear{{Asplund}, {Grevesse}, {Sauval}  \&
  {Scott}}{{Asplund} et~al.}{2009}]{Asplund:2009}
{Asplund} M.,  {Grevesse} N.,  {Sauval} A.~J.,   {Scott} P.,  2009, \mn@doi
  [\araa] {10.1146/annurev.astro.46.060407.145222}, \href
  {http://adsabs.harvard.edu/abs/2009ARA%26A..47..481A} {47, 481}

\bibitem[\protect\citeauthoryear{{Banerjee} et~al.,}{{Banerjee}
  et~al.}{2018}]{Banerjee2018}
{Banerjee} D.~P.~K.,  et~al., 2018, \mn@doi [\apj] {10.3847/1538-4357/aae5d3},
  \href {https://ui.adsabs.harvard.edu/abs/2018ApJ...867...99B} {867, 99}

\bibitem[\protect\citeauthoryear{{Barrett}, {Gaebel}, {Neijssel},
  {Vigna-G{\'o}mez}, {Stevenson}, {Berry}, {Farr}  \& {Mandel}}{{Barrett}
  et~al.}{2018}]{Barrett:2017FIM}
{Barrett} J.~W.,  {Gaebel} S.~M.,  {Neijssel} C.~J.,  {Vigna-G{\'o}mez} A.,
  {Stevenson} S.,  {Berry} C.~P.~L.,  {Farr} W.~M.,   {Mandel} I.,  2018,
  \mn@doi [\mnras] {10.1093/mnras/sty908}, \href
  {http://adsabs.harvard.edu/abs/2018MNRAS.477.4685B} {477, 4685}

\bibitem[\protect\citeauthoryear{Belczynski, Kalogera, Rasio, Taam  \&
  Bulik}{Belczynski et~al.}{2007}]{Belczynski:2006zi}
Belczynski K.,  Kalogera V.,  Rasio F.~A.,  Taam R.~E.,   Bulik T.,  2007,
  \mn@doi [Astrophys. J.] {10.1086/513562}, 662, 504

\bibitem[\protect\citeauthoryear{Belczynski, Holz, Bulik  \&
  O'Shaughnessy}{Belczynski et~al.}{2016}]{Belczynski:2016obo}
Belczynski K.,  Holz D.~E.,  Bulik T.,   O'Shaughnessy R.,  2016, \mn@doi
  [Nature] {10.1038/nature18322}, 534, 512

\bibitem[\protect\citeauthoryear{{Bellm} et~al.,}{{Bellm}
  et~al.}{2019}]{Bellm:2019PASP}
{Bellm} E.~C.,  et~al., 2019, \mn@doi [\pasp] {10.1088/1538-3873/aaecbe}, \href
  {https://ui.adsabs.harvard.edu/abs/2019PASP..131a8002B} {131, 018002}

\bibitem[\protect\citeauthoryear{{Berger} et~al.,}{{Berger}
  et~al.}{2009}]{Berger2009}
{Berger} E.,  et~al., 2009, \mn@doi [\apj] {10.1088/0004-637X/699/2/1850},
  \href {http://adsabs.harvard.edu/abs/2009ApJ...699.1850B} {699, 1850}

\bibitem[\protect\citeauthoryear{{Blagorodnova} et~al.,}{{Blagorodnova}
  et~al.}{2017}]{Blagorodnova:2017}
{Blagorodnova} N.,  et~al., 2017, \mn@doi [\apj] {10.3847/1538-4357/834/2/107},
  \href {http://adsabs.harvard.edu/abs/2017ApJ...834..107B} {834, 107}

\bibitem[\protect\citeauthoryear{{Bond}, {Bedin}, {Bonanos}, {Humphreys},
  {Monard}, {Prieto}  \& {Walter}}{{Bond} et~al.}{2009}]{Bond2009}
{Bond} H.~E.,  {Bedin} L.~R.,  {Bonanos} A.~Z.,  {Humphreys} R.~M.,  {Monard}
  L.~A.~G.~B.,  {Prieto} J.~L.,   {Walter} F.~M.,  2009, \mn@doi [\apjl]
  {10.1088/0004-637X/695/2/L154}, \href
  {http://adsabs.harvard.edu/abs/2009ApJ...695L.154B} {695, L154}

\bibitem[\protect\citeauthoryear{{Cai} et~al.,}{{Cai} et~al.}{2019}]{Cai2019}
{Cai} Y.~Z.,  et~al., 2019, \mn@doi [\aap] {10.1051/0004-6361/201936749}, \href
  {https://ui.adsabs.harvard.edu/abs/2019A&A...632L...6C} {632, L6}

\bibitem[\protect\citeauthoryear{{Chruslinska} \& {Nelemans}}{{Chruslinska} \&
  {Nelemans}}{2019}]{Chruslinska:2019}
{Chruslinska} M.,  {Nelemans} G.,  2019, \mn@doi [\mnras]
  {10.1093/mnras/stz2057}, \href
  {https://ui.adsabs.harvard.edu/abs/2019MNRAS.488.5300C} {488, 5300}

\bibitem[\protect\citeauthoryear{{Clayton}, {Podsiadlowski}, {Ivanova}  \&
  {Justham}}{{Clayton} et~al.}{2017}]{Clayton:2017}
{Clayton} M.,  {Podsiadlowski} P.,  {Ivanova} N.,   {Justham} S.,  2017,
  \mn@doi [\mnras] {10.1093/mnras/stx1290}, \href
  {http://adsabs.harvard.edu/abs/2017MNRAS.470.1788C} {470, 1788}

\bibitem[\protect\citeauthoryear{{De Marco}, {Passy}, {Moe}, {Herwig}, {Mac
  Low}  \& {Paxton}}{{De Marco} et~al.}{2011}]{DeMarco2011}
{De Marco} O.,  {Passy} J.-C.,  {Moe} M.,  {Herwig} F.,  {Mac Low} M.-M.,
  {Paxton} B.,  2011, \mn@doi [\mnras] {10.1111/j.1365-2966.2010.17891.x},
  \href {http://adsabs.harvard.edu/abs/2011MNRAS.411.2277D} {411, 2277}

\bibitem[\protect\citeauthoryear{{Dewi} \& {Tauris}}{{Dewi} \&
  {Tauris}}{2000}]{DewiTauris:2000}
{Dewi} J.~D.~M.,  {Tauris} T.~M.,  2000, Astron. Astrophys., \href
  {http://adsabs.harvard.edu/abs/2000A%26A...360.1043D} {360, 1043}

\bibitem[\protect\citeauthoryear{Dominik, Belczynski, Fryer, Holz, Berti,
  Bulik, Mandel  \& O'Shaughnessy}{Dominik et~al.}{2012}]{Dominik:2012kk}
Dominik M.,  Belczynski K.,  Fryer C.,  Holz D.,  Berti E.,  Bulik T.,  Mandel
  I.,   O'Shaughnessy R.,  2012, \mn@doi [Astrophys. J.]
  {10.1088/0004-637X/759/1/52}, 759, 52

\bibitem[\protect\citeauthoryear{{Eggleton}}{{Eggleton}}{1983}]{Eggleton1983rocheLobe}
{Eggleton} P.~P.,  1983, \mn@doi [\apj] {10.1086/160960}, \href
  {http://adsabs.harvard.edu/abs/1983ApJ...268..368E} {268, 368}

\bibitem[\protect\citeauthoryear{{Ekstr{\"o}m} et~al.,}{{Ekstr{\"o}m}
  et~al.}{2012}]{2012A&A...537A.146E}
{Ekstr{\"o}m} S.,  et~al., 2012, \mn@doi [\aap] {10.1051/0004-6361/201117751},
  \href {https://ui.adsabs.harvard.edu/abs/2012A&A...537A.146E} {537, A146}

\bibitem[\protect\citeauthoryear{{Farr}, {Gair}, {Mandel}  \& {Cutler}}{{Farr}
  et~al.}{2015}]{Farr:2015}
{Farr} W.~M.,  {Gair} J.~R.,  {Mandel} I.,   {Cutler} C.,  2015, \mn@doi [\prd]
  {10.1103/PhysRevD.91.023005}, \href
  {http://adsabs.harvard.edu/abs/2015PhRvD..91b3005F} {91, 023005}

\bibitem[\protect\citeauthoryear{{Ge}, {Webbink}, {Chen}  \& {Han}}{{Ge}
  et~al.}{2015}]{ge2015adiabatic}
{Ge} H.,  {Webbink} R.~F.,  {Chen} X.,   {Han} Z.,  2015, \mn@doi [\apj]
  {10.1088/0004-637X/812/1/40}, \href
  {http://adsabs.harvard.edu/abs/2015ApJ...812...40G} {812, 40}

\bibitem[\protect\citeauthoryear{{Glebbeek}, {Gaburov}, {Portegies Zwart}  \&
  {Pols}}{{Glebbeek} et~al.}{2013}]{Glebbeek2013}
{Glebbeek} E.,  {Gaburov} E.,  {Portegies Zwart} S.,   {Pols} O.~R.,  2013,
  \mn@doi [\mnras] {10.1093/mnras/stt1268}, \href
  {http://adsabs.harvard.edu/abs/2013MNRAS.434.3497G} {434, 3497}

\bibitem[\protect\citeauthoryear{{Goranskij} et~al.,}{{Goranskij}
  et~al.}{2016}]{Goranski2016}
{Goranskij} V.~P.,  et~al., 2016, \mn@doi [Astrophysical Bulletin]
  {10.1134/S1990341316010090}, \href
  {http://adsabs.harvard.edu/abs/2016AstBu..71...82G} {71, 82}

\bibitem[\protect\citeauthoryear{{Graham} et~al.,}{{Graham}
  et~al.}{2019}]{Graham:2019PASP}
{Graham} M.~J.,  et~al., 2019, \mn@doi [\pasp] {10.1088/1538-3873/ab006c},
  \href {https://ui.adsabs.harvard.edu/abs/2019PASP..131g8001G} {131, 078001}

\bibitem[\protect\citeauthoryear{{Groh} et~al.,}{{Groh}
  et~al.}{2019}]{Groh:2019}
{Groh} J.~H.,  et~al., 2019, \mn@doi [\aap] {10.1051/0004-6361/201833720},
  \href {https://ui.adsabs.harvard.edu/abs/2019A&A...627A..24G} {627, A24}

\bibitem[\protect\citeauthoryear{{Hayashi} \& {Nakano}}{{Hayashi} \&
  {Nakano}}{1963}]{1963PThPh..30..460H}
{Hayashi} C.,  {Nakano} T.,  1963, \mn@doi [Progress of Theoretical Physics]
  {10.1143/PTP.30.460}, \href
  {http://adsabs.harvard.edu/abs/1963PThPh..30..460H} {30, 460}

\bibitem[\protect\citeauthoryear{{Hjellming} \& {Webbink}}{{Hjellming} \&
  {Webbink}}{1987}]{Hjellming:1987}
{Hjellming} M.~S.,  {Webbink} R.~F.,  1987, \mn@doi [\apj] {10.1086/165412},
  \href {http://adsabs.harvard.edu/abs/1987ApJ...318..794H} {318, 794}

\bibitem[\protect\citeauthoryear{{Humphreys} \& {Davidson}}{{Humphreys} \&
  {Davidson}}{1979}]{1979ApJ...232..409H}
{Humphreys} R.~M.,  {Davidson} K.,  1979, \mn@doi [\apj] {10.1086/157301},
  \href {https://ui.adsabs.harvard.edu/abs/1979ApJ...232..409H} {232, 409}

\bibitem[\protect\citeauthoryear{{Hurley}, {Pols}  \& {Tout}}{{Hurley}
  et~al.}{2000}]{Hurley2000}
{Hurley} J.~R.,  {Pols} O.~R.,   {Tout} C.~A.,  2000, \mn@doi [\mnras]
  {10.1046/j.1365-8711.2000.03426.x}, \href
  {http://adsabs.harvard.edu/abs/2000MNRAS.315..543H} {315, 543}

\bibitem[\protect\citeauthoryear{{Iaconi} \& {De Marco}}{{Iaconi} \& {De
  Marco}}{2019}]{Iaconi2019}
{Iaconi} R.,  {De Marco} O.,  2019, \mn@doi [\mnras] {10.1093/mnras/stz2756},
  \href {https://ui.adsabs.harvard.edu/abs/2019MNRAS.490.2550I} {490, 2550}

\bibitem[\protect\citeauthoryear{{Iben} \& {Tutukov}}{{Iben} \&
  {Tutukov}}{1984}]{Iben:1984}
{Iben} Jr. I.,  {Tutukov} A.~V.,  1984, \mn@doi [Astrophys. J. Suppl.]
  {10.1086/190932}, \href {http://adsabs.harvard.edu/abs/1984ApJS...54..335I}
  {54, 335}

\bibitem[\protect\citeauthoryear{{Ivanova}}{{Ivanova}}{2011}]{Ivanova2011b}
{Ivanova} N.,  2011, \mn@doi [\apj] {10.1088/0004-637X/730/2/76}, \href
  {https://ui.adsabs.harvard.edu/abs/2011ApJ...730...76I} {730, 76}

\bibitem[\protect\citeauthoryear{{Ivanova} \& {Chaichenets}}{{Ivanova} \&
  {Chaichenets}}{2011}]{Ivanova2011a}
{Ivanova} N.,  {Chaichenets} S.,  2011, \mn@doi [\apjl]
  {10.1088/2041-8205/731/2/L36}, \href
  {http://adsabs.harvard.edu/abs/2011ApJ...731L..36I} {731, L36}

\bibitem[\protect\citeauthoryear{{Ivanova} et~al.,}{{Ivanova}
  et~al.}{2013a}]{Ivanova:2012vx}
{Ivanova} N.,  et~al., 2013a, \mn@doi [\aapr] {10.1007/s00159-013-0059-2},
  \href {http://adsabs.harvard.edu/abs/2013A%26ARv..21...59I} {21, 59}

\bibitem[\protect\citeauthoryear{Ivanova, Justham, Nandez  \& Lombardi}{Ivanova
  et~al.}{2013b}]{Ivanova:2013db}
Ivanova N.,  Justham S.,  Nandez J. L.~A.,   Lombardi Jr J.~C.,  2013b, \mn@doi
  [Science] {10.1126/science.1225540}, 339, 433

\bibitem[\protect\citeauthoryear{{Jencson} et~al.,}{{Jencson}
  et~al.}{2019}]{Jencson:2019ApJL}
{Jencson} J.~E.,  et~al., 2019, \mn@doi [\apjl] {10.3847/2041-8213/ab2c05},
  \href {https://ui.adsabs.harvard.edu/abs/2019ApJ...880L..20J} {880, L20}

\bibitem[\protect\citeauthoryear{{Kami{\'n}ski}, {Menten}, {Tylenda}, {Hajduk},
  {Patel}  \& {Kraus}}{{Kami{\'n}ski} et~al.}{2015}]{2015Natur.520..322K}
{Kami{\'n}ski} T.,  {Menten} K.~M.,  {Tylenda} R.,  {Hajduk} M.,  {Patel}
  N.~A.,   {Kraus} A.,  2015, \mn@doi [\nat] {10.1038/nature14257}, \href
  {http://adsabs.harvard.edu/abs/2015Natur.520..322K} {520, 322}

\bibitem[\protect\citeauthoryear{{Kankare} et~al.,}{{Kankare}
  et~al.}{2015}]{Kankare2015}
{Kankare} E.,  et~al., 2015, \mn@doi [\aap] {10.1051/0004-6361/201526631},
  \href {https://ui.adsabs.harvard.edu/abs/2015A&A...581L...4K} {581, L4}

\bibitem[\protect\citeauthoryear{{Kasen} \& {Woosley}}{{Kasen} \&
  {Woosley}}{2009}]{Kasen2009}
{Kasen} D.,  {Woosley} S.~E.,  2009, \mn@doi [\apj]
  {10.1088/0004-637X/703/2/2205}, \href
  {http://adsabs.harvard.edu/abs/2009ApJ...703.2205K} {703, 2205}

\bibitem[\protect\citeauthoryear{{Kasliwal} et~al.,}{{Kasliwal}
  et~al.}{2011}]{Kasliwal2011}
{Kasliwal} M.~M.,  et~al., 2011, \mn@doi [\apj] {10.1088/0004-637X/730/2/134},
  \href {http://adsabs.harvard.edu/abs/2011ApJ...730..134K} {730, 134}

\bibitem[\protect\citeauthoryear{Kasliwal et~al.}{Kasliwal
  et~al.}{2017}]{Kasliwal:2017fqr}
Kasliwal M.~M.,  et~al., 2017, \mn@doi [Astrophys. J.]
  {10.3847/1538-4357/aa6978}, 839, 88

\bibitem[\protect\citeauthoryear{Kimeswenger}{Kimeswenger}{2006}]{Kimeswenger:2005bva}
Kimeswenger S.,  2006, \mn@doi [Astron. Nachr.] {10.1002/asna.200510482}, 327,
  44

\bibitem[\protect\citeauthoryear{{Klencki}, {Moe}, {Gladysz}, {Chruslinska},
  {Holz}  \& {Belczynski}}{{Klencki} et~al.}{2018}]{Klencki:2018}
{Klencki} J.,  {Moe} M.,  {Gladysz} W.,  {Chruslinska} M.,  {Holz} D.~E.,
  {Belczynski} K.,  2018, \mn@doi [\aap] {10.1051/0004-6361/201833025}, \href
  {https://ui.adsabs.harvard.edu/abs/2018A&A...619A..77K} {619, A77}

\bibitem[\protect\citeauthoryear{{Kochanek}, {Adams}  \&
  {Belczynski}}{{Kochanek} et~al.}{2014}]{Kochanek2014}
{Kochanek} C.~S.,  {Adams} S.~M.,   {Belczynski} K.,  2014, \mn@doi [\mnras]
  {10.1093/mnras/stu1226}, \href
  {http://adsabs.harvard.edu/abs/2014MNRAS.443.1319K} {443, 1319}

\bibitem[\protect\citeauthoryear{Kroupa}{Kroupa}{2001}]{Kroupa:2000iv}
Kroupa P.,  2001, \mn@doi [Mon. Not. Roy. Astron. Soc.]
  {10.1046/j.1365-8711.2001.04022.x}, 322, 231

\bibitem[\protect\citeauthoryear{Kruckow, Tauris, Langer, Szécsi, Marchant  \&
  Podsiadlowski}{Kruckow et~al.}{2016}]{Kruckow2016}
Kruckow M.~U.,  Tauris T.~M.,  Langer N.,  Szécsi D.,  Marchant P.,
  Podsiadlowski P.,  2016, \mn@doi [Astron. Astrophys.]
  {10.1051/0004-6361/201629420}, 596, A58

\bibitem[\protect\citeauthoryear{Kulkarni et~al.}{Kulkarni
  et~al.}{2007}]{Kulkarni:2007kf}
Kulkarni S.~R.,  et~al., 2007, \mn@doi [Nature] {10.1038/nature05822}, 447, 458

\bibitem[\protect\citeauthoryear{{Kulkarni}, {Perley}  \& {Miller}}{{Kulkarni}
  et~al.}{2018}]{Kulkarni2018}
{Kulkarni} S.~R.,  {Perley} D.~A.,   {Miller} A.~A.,  2018, \mn@doi [\apj]
  {10.3847/1538-4357/aabf85}, \href
  {https://ui.adsabs.harvard.edu/abs/2018ApJ...860...22K} {860, 22}

\bibitem[\protect\citeauthoryear{{Kumar}}{{Kumar}}{1963}]{1963ApJ...137.1121K}
{Kumar} S.~S.,  1963, \mn@doi [\apj] {10.1086/147589}, \href
  {http://adsabs.harvard.edu/abs/1963ApJ...137.1121K} {137, 1121}

\bibitem[\protect\citeauthoryear{{Kurtenkov} et~al.,}{{Kurtenkov}
  et~al.}{2015}]{Kurtenkov:2015}
{Kurtenkov} A.~A.,  et~al., 2015, \mn@doi [\aap] {10.1051/0004-6361/201526564},
  \href {http://adsabs.harvard.edu/abs/2015A%26A...578L..10K} {578, L10}

\bibitem[\protect\citeauthoryear{{LSST Science Collaboration} et~al.,}{{LSST
  Science Collaboration} et~al.}{2009}]{LSSTScience:2009}
{LSST Science Collaboration} et~al., 2009, preprint, \href
  {http://adsabs.harvard.edu/abs/2009arXiv0912.0201L} {} (\mn@eprint {arXiv}
  {0912.0201})

\bibitem[\protect\citeauthoryear{{LSST Science Collaboration} et~al.,}{{LSST
  Science Collaboration} et~al.}{2017}]{LSST2017}
{LSST Science Collaboration} et~al., 2017, arXiv e-prints, \href
  {https://ui.adsabs.harvard.edu/abs/2017arXiv170804058L} {p. arXiv:1708.04058}

\bibitem[\protect\citeauthoryear{{Levesque}}{{Levesque}}{2017}]{Levesque:2017}
{Levesque} E.~M.,  2017, {Astrophysics of Red Supergiants},
  \mn@doi{10.1088/978-0-7503-1329-2.
}

\bibitem[\protect\citeauthoryear{{Licquia} \& {Newman}}{{Licquia} \&
  {Newman}}{2015}]{Licquia2015a}
{Licquia} T.~C.,  {Newman} J.~A.,  2015, \mn@doi [\apj]
  {10.1088/0004-637X/806/1/96}, \href
  {https://ui.adsabs.harvard.edu/abs/2015ApJ...806...96L} {806, 96}

\bibitem[\protect\citeauthoryear{{Licquia}, {Newman}  \&
  {Brinchmann}}{{Licquia} et~al.}{2015}]{Licquia2015b}
{Licquia} T.~C.,  {Newman} J.~A.,   {Brinchmann} J.,  2015, \mn@doi [\apj]
  {10.1088/0004-637X/809/1/96}, \href
  {https://ui.adsabs.harvard.edu/abs/2015ApJ...809...96L} {809, 96}

\bibitem[\protect\citeauthoryear{{Lipunov} et~al.,}{{Lipunov}
  et~al.}{2017}]{Lipunov2017}
{Lipunov} V.~M.,  et~al., 2017, \mn@doi [\mnras] {10.1093/mnras/stx1107}, \href
  {https://ui.adsabs.harvard.edu/abs/2017MNRAS.470.2339L} {470, 2339}

\bibitem[\protect\citeauthoryear{{Lombardi}, {Warren}, {Rasio}, {Sills}  \&
  {Warren}}{{Lombardi} et~al.}{2002}]{Lombardi2002}
{Lombardi} Jr. J.~C.,  {Warren} J.~S.,  {Rasio} F.~A.,  {Sills} A.,   {Warren}
  A.~R.,  2002, \mn@doi [\apj] {10.1086/339060}, \href
  {http://adsabs.harvard.edu/abs/2002ApJ...568..939L} {568, 939}

\bibitem[\protect\citeauthoryear{{Loveridge}, {van der Sluys}  \&
  {Kalogera}}{{Loveridge} et~al.}{2011}]{Loveridge2011}
{Loveridge} A.~J.,  {van der Sluys} M.~V.,   {Kalogera} V.,  2011, \mn@doi
  [\apj] {10.1088/0004-637X/743/1/49}, \href
  {http://adsabs.harvard.edu/abs/2011ApJ...743...49L} {743, 49}

\bibitem[\protect\citeauthoryear{{MacLeod}, {Macias}, {Ramirez-Ruiz},
  {Grindlay}, {Batta}  \& {Montes}}{{MacLeod} et~al.}{2017}]{Macleod2017}
{MacLeod} M.,  {Macias} P.,  {Ramirez-Ruiz} E.,  {Grindlay} J.,  {Batta} A.,
  {Montes} G.,  2017, \mn@doi [\apj] {10.3847/1538-4357/835/2/282}, \href
  {http://adsabs.harvard.edu/abs/2017ApJ...835..282M} {835, 282}

\bibitem[\protect\citeauthoryear{{Madau} \& {Dickinson}}{{Madau} \&
  {Dickinson}}{2014}]{Madau2014}
{Madau} P.,  {Dickinson} M.,  2014, \mn@doi [\araa]
  {10.1146/annurev-astro-081811-125615}, \href
  {http://adsabs.harvard.edu/abs/2014ARA%26A..52..415M} {52, 415}

\bibitem[\protect\citeauthoryear{{Martini}, {Wagner}, {Tomaney}, {Rich}, {della
  Valle}  \& {Hauschildt}}{{Martini} et~al.}{1999}]{Martini1999}
{Martini} P.,  {Wagner} R.~M.,  {Tomaney} A.,  {Rich} R.~M.,  {della Valle} M.,
    {Hauschildt} P.~H.,  1999, \mn@doi [\aj] {10.1086/300951}, \href
  {http://adsabs.harvard.edu/abs/1999AJ....118.1034M} {118, 1034}

\bibitem[\protect\citeauthoryear{{Mauerhan}, {Van Dyk}, {Johansson}, {Fox},
  {Filippenko}  \& {Graham}}{{Mauerhan} et~al.}{2018}]{Mauerhan2018}
{Mauerhan} J.~C.,  {Van Dyk} S.~D.,  {Johansson} J.,  {Fox} O.~D.,
  {Filippenko} A.~V.,   {Graham} M.~L.,  2018, \mn@doi [\mnras]
  {10.1093/mnras/stx2500}, \href
  {https://ui.adsabs.harvard.edu/abs/2018MNRAS.473.3765M} {473, 3765}

\bibitem[\protect\citeauthoryear{{Metzger} \& {Pejcha}}{{Metzger} \&
  {Pejcha}}{2017}]{Metzger:2017wrz}
{Metzger} B.~D.,  {Pejcha} O.,  2017, \mn@doi [\mnras] {10.1093/mnras/stx1768},
  \href {http://adsabs.harvard.edu/abs/2017MNRAS.471.3200M} {471, 3200}

\bibitem[\protect\citeauthoryear{{Miller} \& {Scalo}}{{Miller} \&
  {Scalo}}{1979}]{1979ApJS...41..513M}
{Miller} G.~E.,  {Scalo} J.~M.,  1979, \mn@doi [\apjs] {10.1086/190629}, \href
  {https://ui.adsabs.harvard.edu/abs/1979ApJS...41..513M} {41, 513}

\bibitem[\protect\citeauthoryear{{Minniti} et~al.,}{{Minniti}
  et~al.}{2017}]{Minniti2017}
{Minniti} D.,  et~al., 2017, \mn@doi [\apjl] {10.3847/2041-8213/aa9374}, \href
  {http://adsabs.harvard.edu/abs/2017ApJ...849L..23M} {849, L23}

\bibitem[\protect\citeauthoryear{{Moe} \& {Di Stefano}}{{Moe} \& {Di
  Stefano}}{2017}]{Moe:2017ApJS}
{Moe} M.,  {Di Stefano} R.,  2017, \mn@doi [The Astrophysical Journal
  Supplement Series] {10.3847/1538-4365/aa6fb6}, \href
  {https://ui.adsabs.harvard.edu/abs/2017ApJS..230...15M} {230, 15}

\bibitem[\protect\citeauthoryear{{Mould} et~al.,}{{Mould}
  et~al.}{1990}]{Mould1990}
{Mould} J.,  et~al., 1990, \mn@doi [\apjl] {10.1086/185702}, \href
  {http://adsabs.harvard.edu/abs/1990ApJ...353L..35M} {353, L35}

\bibitem[\protect\citeauthoryear{{Nandez} \& {Ivanova}}{{Nandez} \&
  {Ivanova}}{2016}]{Nandez2016}
{Nandez} J.~L.~A.,  {Ivanova} N.,  2016, \mn@doi [\mnras]
  {10.1093/mnras/stw1266}, \href
  {http://adsabs.harvard.edu/abs/2016MNRAS.460.3992N} {460, 3992}

\bibitem[\protect\citeauthoryear{{Nandez}, {Ivanova}  \& {Lombardi}}{{Nandez}
  et~al.}{2014}]{Nandez2014}
{Nandez} J.~L.~A.,  {Ivanova} N.,   {Lombardi} Jr. J.~C.,  2014, \mn@doi [\apj]
  {10.1088/0004-637X/786/1/39}, \href
  {http://adsabs.harvard.edu/abs/2014ApJ...786...39N} {786, 39}

\bibitem[\protect\citeauthoryear{{Nariai} \& {Sugimoto}}{{Nariai} \&
  {Sugimoto}}{1976}]{Nariai:1976}
{Nariai} K.,  {Sugimoto} D.,  1976, \pasj, \href
  {https://ui.adsabs.harvard.edu/abs/1976PASJ...28..593N} {28, 593}

\bibitem[\protect\citeauthoryear{{Neijssel} et~al.,}{{Neijssel}
  et~al.}{2019}]{Neijssel:2019}
{Neijssel} C.~J.,  et~al., 2019, \mn@doi [\mnras] {10.1093/mnras/stz2840},
  \href {https://ui.adsabs.harvard.edu/abs/2019MNRAS.490.3740N} {490, 3740}

\bibitem[\protect\citeauthoryear{Ofek et~al.}{Ofek et~al.}{2008}]{Ofek:2007nw}
Ofek E.~O.,  et~al., 2008, \mn@doi [Astrophys. J.] {10.1086/524350}, 674, 447

\bibitem[\protect\citeauthoryear{{Oskinova}, {Bulik}  \&
  {G{\'o}mez-Mor{\'a}n}}{{Oskinova} et~al.}{2018}]{Oskinova:2018A&AL}
{Oskinova} L.~M.,  {Bulik} T.,   {G{\'o}mez-Mor{\'a}n} A.~N.,  2018, \mn@doi
  [\aap] {10.1051/0004-6361/201832925}, \href
  {https://ui.adsabs.harvard.edu/abs/2018A&A...613L..10O} {613, L10}

\bibitem[\protect\citeauthoryear{{Paczynski}}{{Paczynski}}{1976}]{Paczynski:1976}
{Paczynski} B.,  1976, in {Eggleton} P.,  {Mitton} S.,   {Whelan} J.,  eds,
  IAU Symposium Vol. 73, Structure and Evolution of Close Binary Systems. p.~75

\bibitem[\protect\citeauthoryear{Pastorello et~al.}{Pastorello
  et~al.}{2007}]{Pastorello:2007em}
Pastorello A.,  et~al., 2007, \mn@doi [Nature] {10.1038/nature06282}, 449, E1

\bibitem[\protect\citeauthoryear{{Pastorello} et~al.,}{{Pastorello}
  et~al.}{2019}]{Pastorello2019}
{Pastorello} A.,  et~al., 2019, \mn@doi [\aap] {10.1051/0004-6361/201935999},
  \href {https://ui.adsabs.harvard.edu/abs/2019A&A...630A..75P} {630, A75}

\bibitem[\protect\citeauthoryear{{Pavlovskii}, {Ivanova}, {Belczynski}  \&
  {Van}}{{Pavlovskii} et~al.}{2017}]{Pavlovskii:2016}
{Pavlovskii} K.,  {Ivanova} N.,  {Belczynski} K.,   {Van} K.~X.,  2017, \mn@doi
  [\mnras] {10.1093/mnras/stw2786}, \href
  {https://ui.adsabs.harvard.edu/abs/2017MNRAS.465.2092P} {465, 2092}

\bibitem[\protect\citeauthoryear{{Politano}, {van der Sluys}, {Taam}  \&
  {Willems}}{{Politano} et~al.}{2010}]{Politano:2010}
{Politano} M.,  {van der Sluys} M.,  {Taam} R.~E.,   {Willems} B.,  2010,
  \mn@doi [\apj] {10.1088/0004-637X/720/2/1752}, \href
  {http://adsabs.harvard.edu/abs/2010ApJ...720.1752P} {720, 1752}

\bibitem[\protect\citeauthoryear{{Popov}}{{Popov}}{1993}]{Popov1993}
{Popov} D.~V.,  1993, \mn@doi [\apj] {10.1086/173117}, \href
  {http://adsabs.harvard.edu/abs/1993ApJ...414..712P} {414, 712}

\bibitem[\protect\citeauthoryear{{Prieto} et~al.,}{{Prieto}
  et~al.}{2008}]{Prieto2008}
{Prieto} J.~L.,  et~al., 2008, \mn@doi [\apjl] {10.1086/589922}, \href
  {http://adsabs.harvard.edu/abs/2008ApJ...681L...9P} {681, L9}

\bibitem[\protect\citeauthoryear{{Prieto}, {Sellgren}, {Thompson}  \&
  {Kochanek}}{{Prieto} et~al.}{2009}]{Prieto2009}
{Prieto} J.~L.,  {Sellgren} K.,  {Thompson} T.~A.,   {Kochanek} C.~S.,  2009,
  \mn@doi [\apj] {10.1088/0004-637X/705/2/1425}, \href
  {http://adsabs.harvard.edu/abs/2009ApJ...705.1425P} {705, 1425}

\bibitem[\protect\citeauthoryear{{Raghavan} et~al.,}{{Raghavan}
  et~al.}{2010}]{Raghavan:2010ApJS}
{Raghavan} D.,  et~al., 2010, \mn@doi [The Astrophysical Journal Supplement
  Series] {10.1088/0067-0049/190/1/1}, \href
  {https://ui.adsabs.harvard.edu/\#abs/2010ApJS..190....1R} {190, 1}

\bibitem[\protect\citeauthoryear{Rau, Kulkarni, Ofek  \& Yan}{Rau
  et~al.}{2007}]{Rau:2006vc}
Rau A.,  Kulkarni S.~R.,  Ofek E.~O.,   Yan L.,  2007, \mn@doi [Astrophys. J.]
  {10.1086/512672}, 659, 1536

\bibitem[\protect\citeauthoryear{{Salpeter}}{{Salpeter}}{1955}]{Salpeter:1955ApJ}
{Salpeter} E.~E.,  1955, \mn@doi [\apj] {10.1086/145971}, \href
  {http://adsabs.harvard.edu/abs/1955ApJ...121..161S} {121, 161}

\bibitem[\protect\citeauthoryear{{Sana} et~al.,}{{Sana}
  et~al.}{2012}]{Sana:2012}
{Sana} H.,  et~al., 2012, \mn@doi [Science] {10.1126/science.1223344}, \href
  {http://adsabs.harvard.edu/abs/2012Sci...337..444S} {337, 444}

\bibitem[\protect\citeauthoryear{{Sanyal}, {Langer}, {Sz{\'e}csi}, {-C Yoon}
  \& {Grassitelli}}{{Sanyal} et~al.}{2017}]{Sanyal:2017}
{Sanyal} D.,  {Langer} N.,  {Sz{\'e}csi} D.,  {-C Yoon} S.,   {Grassitelli} L.,
   2017, \mn@doi [\aap] {10.1051/0004-6361/201629612}, \href
  {https://ui.adsabs.harvard.edu/abs/2017A&A...597A..71S} {597, A71}

\bibitem[\protect\citeauthoryear{{Schr{\o}der}, {MacLeod}, {Loeb},
  {Vigna-G{\'o}mez}  \& {Mandel}}{{Schr{\o}der} et~al.}{2019}]{Schroder:2019}
{Schr{\o}der} S.~L.,  {MacLeod} M.,  {Loeb} A.,  {Vigna-G{\'o}mez} A.,
  {Mandel} I.,  2019, arXiv e-prints, \href
  {https://ui.adsabs.harvard.edu/abs/2019arXiv190604189S} {p. arXiv:1906.04189}

\bibitem[\protect\citeauthoryear{Scott}{Scott}{1992}]{Scott1992}
Scott D.,  1992, Multivariate Density Estimation: Theory, Practice, and
  Visualization.
A Wiley-interscience publication, Wiley, \url
  {https://books.google.com.au/books?id=7crCUS\_F2ocC}

\bibitem[\protect\citeauthoryear{{Segev}, {Sabach}  \& {Soker}}{{Segev}
  et~al.}{2019}]{Segev:2019}
{Segev} R.,  {Sabach} E.,   {Soker} N.,  2019, \mn@doi [\apj]
  {10.3847/1538-4357/ab3f2a}, \href
  {https://ui.adsabs.harvard.edu/abs/2019ApJ...884...58S} {884, 58}

\bibitem[\protect\citeauthoryear{{Smarr} \& {Blandford}}{{Smarr} \&
  {Blandford}}{1976}]{SmarrBlandford:1976}
{Smarr} L.~L.,  {Blandford} R.,  1976, \mn@doi [\apj] {10.1086/154524}, \href
  {http://adsabs.harvard.edu/abs/1976ApJ...207..574S} {207, 574}

\bibitem[\protect\citeauthoryear{Smith et~al.}{Smith
  et~al.}{2016}]{Smith:2016qtr}
Smith N.,  et~al., 2016, \mn@doi [Mon. Not. Roy. Astron. Soc.]
  {10.1093/mnras/stw219}, 458, 950

\bibitem[\protect\citeauthoryear{{Soberman}, {Phinney}  \& {van den
  Heuvel}}{{Soberman} et~al.}{1997}]{soberman1997stability}
{Soberman} G.~E.,  {Phinney} E.~S.,   {van den Heuvel} E.~P.~J.,  1997, \aap,
  \href {http://adsabs.harvard.edu/abs/1997A%26A...327..620S} {327, 620}

\bibitem[\protect\citeauthoryear{{Soker} \& {Tylenda}}{{Soker} \&
  {Tylenda}}{2003}]{SokerTylenda:2003}
{Soker} N.,  {Tylenda} R.,  2003, \mn@doi [\apjl] {10.1086/367759}, \href
  {http://adsabs.harvard.edu/abs/2003ApJ...582L.105S} {582, L105}

\bibitem[\protect\citeauthoryear{{Stevenson}, {Vigna-G{\'o}mez}, {Mandel},
  {Barrett}, {Neijssel}, {Perkins}  \& {de Mink}}{{Stevenson}
  et~al.}{2017}]{Stevenson:2017tfq}
{Stevenson} S.,  {Vigna-G{\'o}mez} A.,  {Mandel} I.,  {Barrett} J.~W.,
  {Neijssel} C.~J.,  {Perkins} D.,   {de Mink} S.~E.,  2017, \mn@doi [Nature
  Communications] {10.1038/ncomms14906}, \href
  {https://ui.adsabs.harvard.edu/abs/2017NatCo...814906S} {8, 14906}

\bibitem[\protect\citeauthoryear{{Stevenson}, {Sampson}, {Powell},
  {Vigna-G{\'o}mez}, {Neijssel}, {Sz{\'e}csi}  \& {Mandel}}{{Stevenson}
  et~al.}{2019}]{Stevenson:2019}
{Stevenson} S.,  {Sampson} M.,  {Powell} J.,  {Vigna-G{\'o}mez} A.,  {Neijssel}
  C.~J.,  {Sz{\'e}csi} D.,   {Mandel} I.,  2019, \mn@doi [\apj]
  {10.3847/1538-4357/ab3981}, \href
  {https://ui.adsabs.harvard.edu/abs/2019ApJ...882..121S} {882, 121}

\bibitem[\protect\citeauthoryear{Tauris \& Dewi}{Tauris \&
  Dewi}{2001}]{Tauris:2001cx}
Tauris T.~M.,  Dewi J. D.~M.,  2001, \mn@doi [Astron. Astrophys.]
  {10.1051/0004-6361:20010099}, 369, 170

\bibitem[\protect\citeauthoryear{{Tout}, {Aarseth}, {Pols}  \&
  {Eggleton}}{{Tout} et~al.}{1997}]{1997MNRAS.291..732T}
{Tout} C.~A.,  {Aarseth} S.~J.,  {Pols} O.~R.,   {Eggleton} P.~P.,  1997,
  \mn@doi [\mnras] {10.1093/mnras/291.4.732}, \href
  {http://adsabs.harvard.edu/abs/1997MNRAS.291..732T} {291, 732}

\bibitem[\protect\citeauthoryear{{Tutukov} \& {Yungelson}}{{Tutukov} \&
  {Yungelson}}{1993}]{Tutukov:1993}
{Tutukov} A.~V.,  {Yungelson} L.~R.,  1993, \mn@doi [\mnras]
  {10.1093/mnras/260.3.675}, \href
  {http://adsabs.harvard.edu/abs/1993MNRAS.260..675T} {260, 675}

\bibitem[\protect\citeauthoryear{{Tylenda}}{{Tylenda}}{2005}]{Tylenda:2005}
{Tylenda} R.,  2005, \mn@doi [\aap] {10.1051/0004-6361:20052800}, \href
  {http://adsabs.harvard.edu/abs/2005A%26A...436.1009T} {436, 1009}

\bibitem[\protect\citeauthoryear{{Tylenda} et~al.,}{{Tylenda}
  et~al.}{2011}]{Tylenda:2011}
{Tylenda} R.,  et~al., 2011, \mn@doi [\aap] {10.1051/0004-6361/201016221},
  \href {http://adsabs.harvard.edu/abs/2011A%26A...528A.114T} {528, A114}

\bibitem[\protect\citeauthoryear{{Tylenda} et~al.,}{{Tylenda}
  et~al.}{2013}]{Tylenda:2013}
{Tylenda} R.,  et~al., 2013, \mn@doi [\aap] {10.1051/0004-6361/201321647},
  \href {http://adsabs.harvard.edu/abs/2013A%26A...555A..16T} {555, A16}

\bibitem[\protect\citeauthoryear{{Van Dyk}, {Peng}, {King}, {Filippenko},
  {Treffers}, {Li}  \& {Richmond}}{{Van Dyk} et~al.}{2000}]{VanDyk2000}
{Van Dyk} S.~D.,  {Peng} C.~Y.,  {King} J.~Y.,  {Filippenko} A.~V.,  {Treffers}
  R.~R.,  {Li} W.,   {Richmond} M.~W.,  2000, \mn@doi [\pasp] {10.1086/317727},
  \href {https://ui.adsabs.harvard.edu/abs/2000PASP..112.1532V} {112, 1532}

\bibitem[\protect\citeauthoryear{{Vigna-G{\'o}mez} et~al.,}{{Vigna-G{\'o}mez}
  et~al.}{2018}]{Vignagomez2018}
{Vigna-G{\'o}mez} A.,  et~al., 2018, \mn@doi [\mnras] {10.1093/mnras/sty2463},
  \href {http://adsabs.harvard.edu/abs/2018MNRAS.481.4009V} {481, 4009}

\bibitem[\protect\citeauthoryear{Voss \& Tauris}{Voss \&
  Tauris}{2003}]{Voss:2003ep}
Voss R.,  Tauris T.~M.,  2003, \mn@doi [Mon. Not. Roy. Astron. Soc.]
  {10.1046/j.1365-8711.2003.06616.x}, 342, 1169

\bibitem[\protect\citeauthoryear{{Wang}, {Jia}  \& {Li}}{{Wang}
  et~al.}{2016}]{Wang2016}
{Wang} C.,  {Jia} K.,   {Li} X.-D.,  2016, \mn@doi [Research in Astronomy and
  Astrophysics] {10.1088/1674-4527/16/8/126}, \href
  {http://adsabs.harvard.edu/abs/2016RAA....16..126W} {16, 126}

\bibitem[\protect\citeauthoryear{{Webbink}}{{Webbink}}{1984}]{Webbink:1984}
{Webbink} R.~F.,  1984, \mn@doi [Astrophys. J.] {10.1086/161701}, \href
  {http://adsabs.harvard.edu/abs/1984ApJ...277..355W} {277, 355}

\bibitem[\protect\citeauthoryear{{Webbink}}{{Webbink}}{2008}]{Webbink:2008}
{Webbink} R.~F.,  2008, {Common Envelope Evolution Redux}.
p.~233 (\mn@eprint {arXiv} {0704.0280}), \mn@doi{10.1007/978-1-4020-6544-6_13}

\bibitem[\protect\citeauthoryear{{White}, {Daw}  \& {Dhillon}}{{White}
  et~al.}{2011}]{White2011}
{White} D.~J.,  {Daw} E.~J.,   {Dhillon} V.~S.,  2011, \mn@doi [Classical and
  Quantum Gravity] {10.1088/0264-9381/28/8/085016}, \href
  {http://adsabs.harvard.edu/abs/2011CQGra..28h5016W} {28, 085016}

\bibitem[\protect\citeauthoryear{{Williams}, {Darnley}, {Bode}  \&
  {Steele}}{{Williams} et~al.}{2015}]{Williams:2015}
{Williams} S.~C.,  {Darnley} M.~J.,  {Bode} M.~F.,   {Steele} I.~A.,  2015,
  \mn@doi [\apjl] {10.1088/2041-8205/805/2/L18}, \href
  {http://adsabs.harvard.edu/abs/2015ApJ...805L..18W} {805, L18}

\bibitem[\protect\citeauthoryear{{Xu} \& {Li}}{{Xu} \& {Li}}{2010a}]{Xu2010}
{Xu} X.-J.,  {Li} X.-D.,  2010a, \mn@doi [\apj] {10.1088/0004-637X/716/1/114},
  \href {http://adsabs.harvard.edu/abs/2010ApJ...716..114X} {716, 114}

\bibitem[\protect\citeauthoryear{{Xu} \& {Li}}{{Xu} \&
  {Li}}{2010b}]{XuLi:2010err}
{Xu} X.-J.,  {Li} X.-D.,  2010b, \mn@doi [\apj] {10.1088/0004-637X/722/2/1985},
  \href {https://ui.adsabs.harvard.edu/abs/2010ApJ...722.1985X} {722, 1985}

\bibitem[\protect\citeauthoryear{{de Kool}, {van den Heuvel}  \& {Pylyser}}{{de
  Kool} et~al.}{1987}]{deKool:1987}
{de Kool} M.,  {van den Heuvel} E.~P.~J.,   {Pylyser} E.,  1987, Astron.
  Astrophys., \href {http://adsabs.harvard.edu/abs/1987A%26A...183...47D} {183,
  47}

\bibitem[\protect\citeauthoryear{de Mink, Sana, Langer, Izzard  \&
  Schneider}{de~Mink et~al.}{2014}]{deMink:2013xqa}
de Mink S.~E.,  Sana H.,  Langer N.,  Izzard R.~G.,   Schneider F. R.~N.,
  2014, \mn@doi [Astrophys. J.] {10.1088/0004-637X/782/1/7}, 782, 7

\bibitem[\protect\citeauthoryear{{van den Heuvel}}{{van den
  Heuvel}}{1976}]{vdH:1976}
{van den Heuvel} E.~P.~J.,  1976, in {Eggleton} P.,  {Mitton} S.,   {Whelan}
  J.,  eds,  IAU Symposium Vol. 73, Structure and Evolution of Close Binary
  Systems. p.~35

\makeatother
\end{thebibliography}
